\documentclass[prb,aps,twocolumn,showpacs,superscriptaddress,floatfix]{revtex4}
\usepackage{epsfig}
\usepackage{epstopdf}
\usepackage{amsmath}
\usepackage{amssymb}
\usepackage{amsfonts}
\usepackage{mathptmx}
\usepackage{eucal}
\usepackage{bm}

\graphicspath{{Pictures1/}}
\begin{document}

\title{Current-phase relations in SIsFS junctions in the vicinity of 0-$\pi$
transition}
\author{S.~V.~Bakurskiy}
\affiliation{Skobeltsyn Institute of Nuclear Physics, Lomonosov Moscow State University
1(2), Leninskie gory, Moscow 119234, Russian Federation}
\affiliation{Moscow Institute of Physics and Technology, Dolgoprudny, Moscow Region,
141700, Russian Federation}
\affiliation{National University of Science and Technology MISIS, 4 Leninsky prosp.,
Moscow, 119049, Russia}
\author{V.~I.~Filippov}
\affiliation{Faculty of Physics, M.V. Lomonosov Moscow State University, 119992 Leninskie
Gory, Moscow, Russia}
\affiliation{All-Russian Research Institute of Automatics n.a. N.L. Dukhov (VNIIA), 127055, Moscow, Russia}
\author{V.~I.~Ruzhickiy}
\affiliation{Faculty of Physics, M.V. Lomonosov Moscow State University, 119992 Leninskie
Gory, Moscow, Russia}
\author{N.~V.~Klenov}
\affiliation{Faculty of Physics, M.V. Lomonosov Moscow State University, 119992 Leninskie
Gory, Moscow, Russia}
\affiliation{Moscow Institute of Physics and Technology, Dolgoprudny, Moscow Region,
141700, Russian Federation}
\affiliation{All-Russian Research Institute of Automatics n.a. N.L. Dukhov (VNIIA), 127055, Moscow, Russia}
\author{I.~I.~Soloviev}
\affiliation{Skobeltsyn Institute of Nuclear Physics, Lomonosov Moscow State University
1(2), Leninskie gory, Moscow 119234, Russian Federation}
\affiliation{Moscow Institute of Physics and Technology, Dolgoprudny, Moscow Region,
141700, Russian Federation}
\author{M.~Yu.~Kupriyanov}
\affiliation{Skobeltsyn Institute of Nuclear Physics, Lomonosov Moscow State University
1(2), Leninskie gory, Moscow 119234, Russian Federation}
\affiliation{Moscow Institute of Physics and Technology, Dolgoprudny, Moscow Region,
141700, Russian Federation}
\affiliation{National University of Science and Technology MISIS, 4 Leninsky prosp.,
Moscow, 119049, Russia}
\author{A.~A.~Golubov}
\affiliation{Moscow Institute of Physics and Technology, Dolgoprudny, Moscow Region,
141700, Russian Federation}
\affiliation{Faculty of Science and Technology and MESA+ Institute for Nanotechnology,
University of Twente, 7500 AE Enschede, The Netherlands}
\date{\today }

\begin{abstract}
We consider the current-phase relation (CPR) in the Josephson junctions with complex
insulator-superconductor-ferromagnetic interlayers in the vicinity of 0-$\pi$
transition. We find a strong impact of the second harmonic on CPR of the junctions. It is shown that the critical current can be kept
constant in the region of 0-pi transition, while the CPR
transforms through multi-valued hysteretic states depending on the relative
values of tunnel transparency and magnetic thickness. Moreover,
CPR in the transition region has multiple branches with
distinct ground states.

\end{abstract}

\maketitle

\section{Introduction \label{Intro}}

The current-phase relation (CPR), $I_{S}(\varphi ),$ between a supercurrent,
$I_{S},$ and a phase difference, $\varphi, $ is the most basic property of a
Josephson junction \cite{Lik, RevG}. It is well-known that CPR in a
superconductor-insulator-superconductor (SIS) type junction has a sinusoidal
shape at arbitrary temperatures. In the superconductor-normal-superconductor
(SNS), superconductor-ferromagnetic-superconductor (SFS) junctions or double
barrier SINIS structures, deviations from this behavior occur at
temperatures much smaller than the critical temperature, $T_C$, of S electrodes, $%
T_{C}$. At the same time, in all these structures, $I_{S}(\varphi )$ is a
single-valued function of $\varphi$, irrespective of transport properties
and geometry of a weak-link region \cite{RevG}.

Previously, it was shown that the situation might be different when the weak
link is formed by a material which is intrinsically superconducting (s) with a
transition temperature lower than that of the S electrodes. In this case, an
increase of the distance between the electrodes may result in the transformation
\cite{LikYak} of $I_{S}(\varphi )$ from single- to multi-valued function of $%
\varphi$. The parameter range for which this transformation takes place,
defines the transition from the Josephson effect to Abrikosov vortex flux
flow in the s film \cite{KLM}.

Recent theoretical \cite{Bakurskiy2} and experimental \cite{Ruppelt} studies
indicated a possible realization of the above mentioned transformations of $%
I_{S}(\varphi )$ in SIsFS structures in the form of instabilities near $0$-$%
\pi$ transition. So far, this new fundamental feature of the Josephson
structures remains unexplored. In this paper we address this problem by
considering the properties of SIsFS junction in the vicinity of $0$-$\pi$
transition taking into account the existence of significant second harmonic
of current-phase relation (CPR) in sFS part of the structure.

We find that the $0$ - $\pi $ transition in SIsFS structures is going through
 distinct states with a discontinuous hysteretic current-phase
relations. Moreover, the protected $0$ and $\pi $ \ states are
found in the system with multiple possible branches of current phase
relation. Finally, we demonstrate, that the $0$-$\pi $ transition can be
realized without changes of the critical current due to transformation of
current-phase relations, which hinders an observation of this transition in
the conventional manner and requires phase sensitive experiments \cite{Frolov}.

The paper is organized as follows. In Section II two theoretical models, microscopic and phenomenological one, are formulated, which describe the CPR in SIsFS structures and the results of these two approaches are compared. Sections III and IV provide analytical and numerical
results for CPR followed from the lumped contacts model. The classification of the physical states in the SIsFS structures is introduced in terms of a number of the ground states and shapes of $I_{S}(\varphi )$ curves.

\begin{figure}[t]
\center{\includegraphics[width=0.80\linewidth]{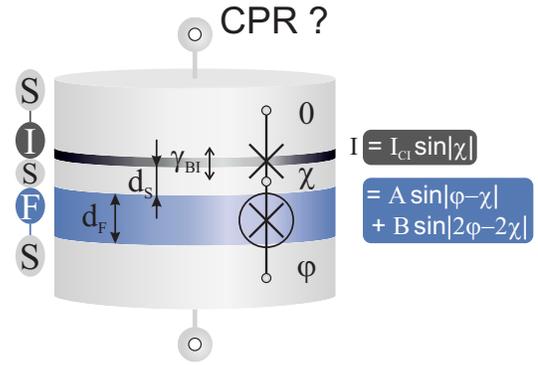}} \vspace{-1 mm}
\caption{Sketch of the SIsFS structure with equivalent scheme for the lumped
elements method.}
\label{CritCurr}
\end{figure}

\section{Theoretical model of SIsFS structure\label{Model}}

Below we will use two complementary approaches for solving the problem. The
first one is based on microscopic theory of superconductivity and employs
numerical simulation of the processes in the structure within the framework of the
Usadel equations \cite{Usadel} with Kupriyanov-Lukichev boundary conditions
\cite{KL} at the interfaces.

\begin{equation}
\frac{\pi T_{C}\xi _{p}^{2}}{\widetilde{\omega }_{p}G_{m}}\frac{d}{dx}\left(
G_{p}^{2}\frac{d\Phi _{p}}{dx}\right) -\Phi _{p}=-\Delta _{p}  \label{fiS1}
\end{equation}%
\begin{equation}
\Delta _{p}\ln \frac{T}{T_{C}}+\frac{T}{T_{C}}\sum_{\omega =-\infty
}^{\infty }\left( \frac{\Delta _{p}}{\left\vert \omega \right\vert }-\frac{%
\Phi _{p}G_{p}}{\omega }\right) =0,  \label{delta}
\end{equation}%
\begin{equation}
\pm \gamma _{Bpq}\xi _{p}G_{p}\frac{d}{dx}\Phi _{p}=G_{q}\left( \frac{%
\widetilde{\omega }_{p}}{\widetilde{\omega }_{q}}\Phi _{q}-\Phi _{p}\right) .
\label{BC}
\end{equation}%
Here $p$ and $q$ are subscripts of corresponding layers, $G_{p}=\widetilde{%
\omega }_{p}/\sqrt{\widetilde{\omega }_{p}^{2}+\Phi _{p,\omega }\Phi
_{p,-\omega }^{\ast }},$ $\widetilde{\omega }_{p}=\omega +iH_{p}$, $\omega
=\pi T(2n+1)$ are the Matsubara frequencies, $\Delta _{p}$ is the pair potential
which exists insode the superconductors, $H_{p},$ is exchange energy of
ferromagnetic layer ($H_{p}=0$ in nonferromagnetic materials), $T_{C},$ is
critical temperature of superconductors, $\xi _{p}=(D_{p}/2\pi T_{C})^{1/2}$
is the coherence length, $D_{p}$ is diffusion coefficient, $G_{p},$ and, $\Phi
_{p},$ are the normal and anomalous Green's functions, respectively, $\gamma
_{Bpq}=R_{Bpq}\mathcal{A}_{Bpq}/\rho _{p}\xi _{p}$, is suppression
parameter, $R_{Bpq}$ and $\mathcal{A}_{Bpq}$ are the resistance and area of
corresponding interface. The sign plus in (\ref{BC}) means that $p$-th
material is located at the side $x_{i}-0$ from interface position $x_{i}$,
and sign minus corresponds to the case than $p$-th material is at $x_{i}+0$.
The $x$ axis is oriented perpendicular to the interfaces. At free surfaces
of the S electrodes located far away from the boundaries $(x\rightarrow \pm
\infty )$ we set bulk values of Green function in superconductor $\Phi
=\Delta _{0}exp(i\psi )$ with $\psi =0$ and $\psi =\varphi$.

The boundary problem (\ref{fiS1}) - (\ref{BC}) was solved numerically by making
use of the algorithm developed in Ref. \cite{Bakurskiy2}. Calculated Green's
functions were used to find a current across a SIsFS junction as a function
of phase difference $\varphi $
\begin{equation}
\frac{2eI_{S}(\varphi )}{\pi T\mathcal{A}_{B}}=\sum\limits_{\omega =-\infty
}^{\infty }\frac{iG_{p,\omega }^{2}}{\rho _{p}\widetilde{\omega }_{p}^{2}}%
\left[ \Phi _{p,\omega }\frac{\partial \Phi _{p,-\omega }^{\ast }}{\partial x%
}-\Phi _{p,-\omega }^{\ast }\frac{\partial \Phi _{p,\omega }}{\partial x}%
\right] .  \label{current}
\end{equation}

The second approach is a phenomenological one. It is based on modelling the
structure as a system of two lumped contacts connected in series (see
Fig. 1): the SIs junction with conventional sinusoidal CPR $I_{SIs}=I_{CI}\sin (\chi )$ and an sFS junction which has CPR
\begin{equation}
I_{sFS}=A\sin (\varphi -\chi )+B\sin (2(\varphi -\chi ))
\label{IsFS}
\end{equation}%
having both the first and the second harmonics. Within this model, the
amplitudes $A$ and $B$ are considered as independent parameters, while the
phase difference on tunnel layer $\chi $ is a function of phase drop on whole junction $\varphi$. The $\chi (\varphi )$ dependence
can be found from the condition of the current equality across SIs and sFS
junctions.

\begin{equation}
I_{CI}\sin (\chi )=A\sin (\varphi -\chi )+B\sin (2(\varphi -\chi )).\
\label{ICeq1}
\end{equation}%
The lumped contacts model is applicable\cite{Bakurskiy2} for $d_{S}>\pi
^{2}\xi _{S}/(4\sqrt{1-T/T_{C}}).$

For arbitrary relations between $I_{CI},$ $A$ and $B$, the equation (\ref%
{ICeq1}) for $\chi (\varphi )$ has been solved numerically, thus determining
the current-phase $I_{S}(\varphi )=I_{CI}\sin (\chi )$ and the energy-phase
relations
\begin{equation}
E(\varphi )=\frac{\Phi _{0}I_{CI}}{2\pi }(1-\cos (\chi ))+E_{A}+E_{B},
\label{EotFi}
\end{equation}%
\begin{equation*}
E_{A}=\frac{\Phi _{0}A}{2\pi }(1-\cos (\varphi -\chi )),\ E_{B}=\frac{\Phi
_{0}B}{4\pi }(1-\cos (2(\varphi -\chi )).
\end{equation*}

Generally, the equation (\ref{ICeq1}) has several independent solutions for $%
\chi (\varphi )$ in the interval $0\leq\varphi \leq 2\pi $. However, only
some of these solutions meet the stability criterion%
\begin{equation}
\frac{d^{2}E(\chi )}{d\chi ^{2}}=\cos (\chi )+\frac{A}{I_{CI}}\cos (\varphi
-\chi )+\frac{2B}{I_{CI}}\cos (2(\varphi -\chi ))>0,  \label{Stab}
\end{equation}%
which means that the solution is stable if the functional $E(\chi )$ for
certain $\varphi $ is at a local minimum.

\begin{figure}[t]
\begin{minipage}[h]{0.95\linewidth}
\center{a)\includegraphics[width=0.95\linewidth]{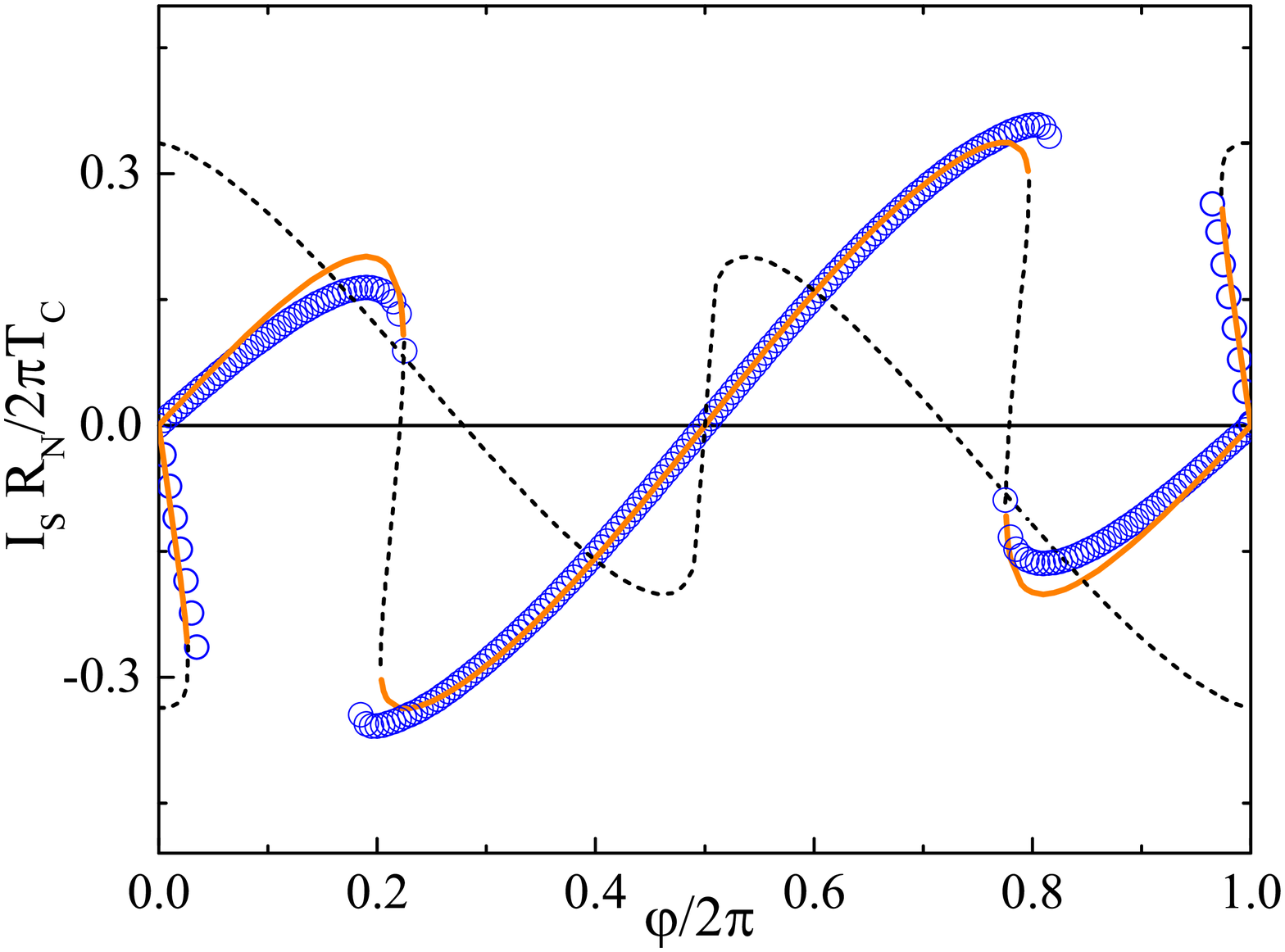}}
\end{minipage}
\vfill
\begin{minipage}[h]{0.95\linewidth}
 \center{b)\includegraphics[width=0.95\linewidth]{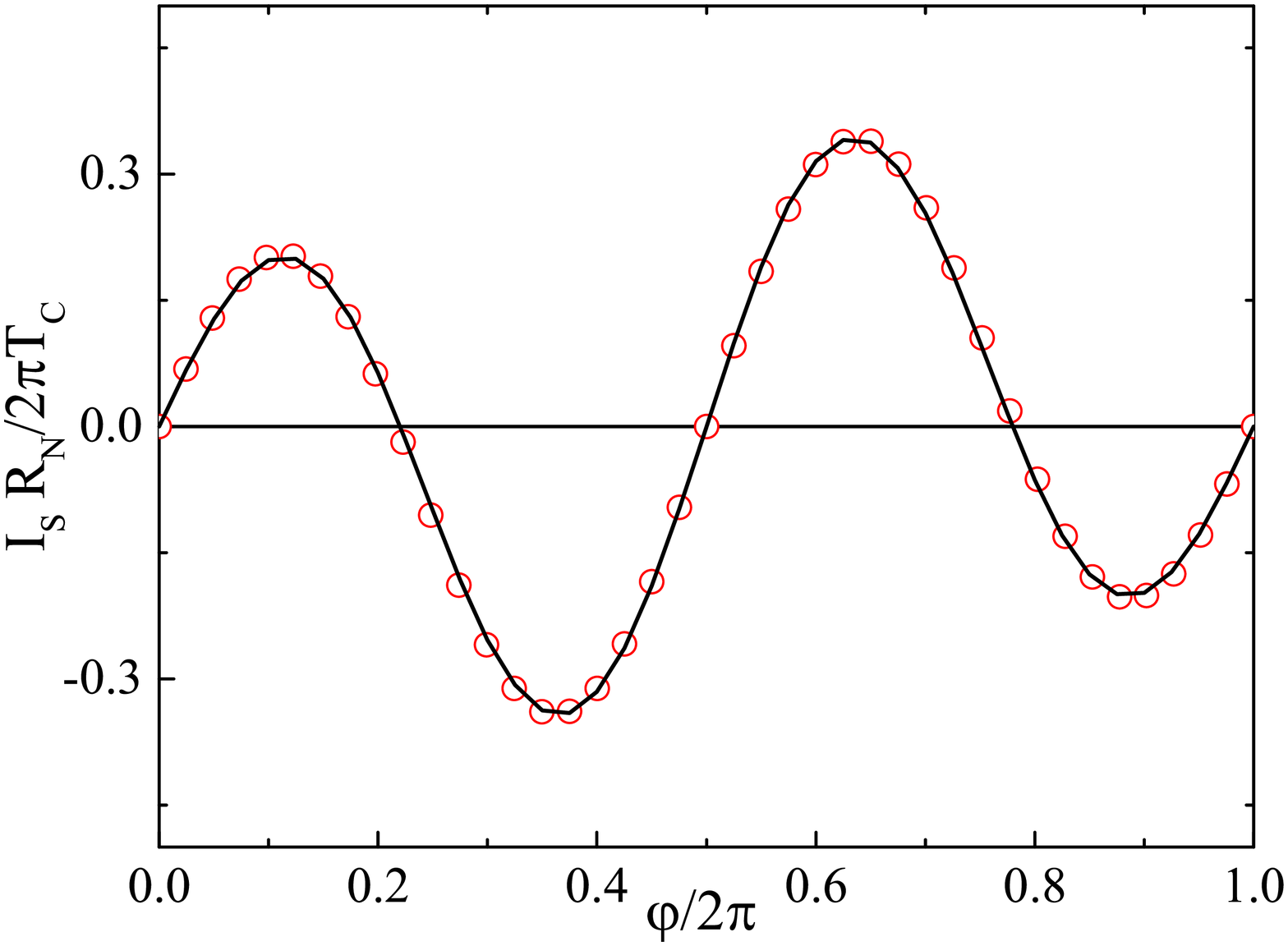}}
\end{minipage}
\caption{(Color Online) The current-phase relation of the SIsFS (panel a)) and
sFS (panel b)) junctions in the vicinity of $0-\protect\pi$ transition.
Panel a) shows comparison between the CPR following from solution of the
Usadel equations (open circles) obtained for $d_{F}=0.46\protect\xi _{S},$ $%
d_{s}=5 \protect\xi _{S},$ $H=10\protect\pi T_{C},$ $T=0.2T_{C}$ and the CPR
calculated in lumped junctions model for $A=-0.22I_{CI}$ and $B=0.61I_{CI}$
(the solid and the dashed lines are stable and unstable parts of solution,
respectively). The open circles in the panel b) give CPR of sFS junction
calculated from the Usadel equations for the same set of the parameters and the
fit is shown with the solid line for $A=-0.22I_{CI}$ and $B=0.61I_{CI}.$ }
\label{compare}
\end{figure}

In the Fig.\ref{compare}a we compare $I_{S}(\varphi )$ dependencies
calculated in the frame of the both approaches. The solution of Usadel equations has
been found for the following set of parameters: $d_{F}=0.46\xi _{S},$ $%
d_{s}=5 \xi _{S},$ $H=10\pi T_{C},$ $T=0.2T_{C},$ the suppression parameters at
SIs and SF interfaces
are equal to $\gamma _{BI}=1000$ and $\gamma _{BSF}=0.3,$ respectively. The
resulting $I_{S}(\varphi )$ dependence of SIsFS\ contact is shown in Fig.\ref%
{compare} by the open circles. It can be seen that there are two critical
points in $I_{S}(\varphi )$ curve at which there is a stepwise change of the
supercurrent. They are located at $\varphi /2\pi \approx 0.2;\ 0.8.$

In the spirit of the lumped junction model, one has to find the
characteristics of SIs and sFS parts of the SIsFS structure independently from
each other. For SIs tunnel junction we get $I_{CI}=0.88\pi T_{C}/R_{N}.$
Microscopic calculations for the sFS structure demonstrate that it
is in a vicinity of the $0-\pi $ transition and its $I_{sFS}(\varphi )$
relationship can be really approximated by Eq. (\ref{IsFS}) with $%
A=-0.22I_{CI}$\thinspace\ and $B=0.61I_{CI}$ (see Fig.\ref{compare}b).

Substitution of this findings into (\ref{ICeq1}) gives the $I_{S}(\varphi )$
presented in Fig.\ref{compare}a by solid and dashed lines, which,
respectively, corresponds to stable and unstable parts of $I_{S}(\varphi )$ curves
calculated in the lumped junctions model for $I_{CI}=0.88\pi T_{C}/R_{N},$ $%
A=-0.22I_{CI}$\thinspace\ and $B=0.61I_{CI}.$ We find a good match between
the shapes of the curves calculated within the framework of these two
approaches. The solutions of the Eq.\ref{ICeq1} shown by the dashed curves on $I_{S}(\varphi )$ dependence
correspond to the local maxima of $E(\chi )$. The system leaves these unstable
states located at $\varphi /2\pi \approx 0.2\ $\ and $\ \varphi /2\pi
\approx 0.8$ through the resistive states of junctions and continuous change of the
phase $\chi $. In the vicinity of $\varphi \approx 2\pi$, the lumped junction
model predicts the existence of two stable branches for the $I_{S}(\varphi )$
dependence. The first one corresponds to the line, with a positive derivative in the vicinity of $\varphi=2\pi$. This branch is stable in the whole range of definition with two breake hysteretic points $\varphi /2\pi \approx 0.2;\ 0.8$. The second branch of  $I_{S}(\varphi )$  has a negative derivative for $\varphi=2\pi$. This solution has stable parts only in the small vicinity of $\varphi=2\pi$, while for the other parameter range, it is unstable ( see the long dashed line on Fig.\ref{compare}a stretching through the whole graph).

The microscopic approach permits to reach both branches depending
on the initial conditions of the iterative calculation process. The
realization of these two stable branches is shown in Fig.\ref{compare}a. The
presence of stable intersecting branches in $I_{S}(\varphi )$ dependence in
the vicinity of $\varphi =2\pi $ is a point for discussion. On the one
hand it leads to the potential instabilities caused by hopping between the
stable states under an influence of external noise environment. While on the
other, the presence of such states is a precondition to the different
applications in logic or memory device. In any case, it is important to study and
classify the variety of such multy-valued states in junctions. To that end, we shall concentrate, hereafter, only on the analysis in the frame of lumped
junctions model. As follows from Fig.\ref{compare}, it may provide all stable
and unstable branches of $I_{S}(\varphi )$ dependence, which fits reasonably
well the exact result obtained from microscopic theory. The latter requires
much longer calculation time; especially, for the thick middle s-layer and low
temperatures due to slow convergence in self-consistent iteration cycle. In
addition, the result of iterative process during solving the microscopic problem
is sensitive to initial parameters, i.e. initial phase of the intermediate
s-electrode.

Finally, without loss of generality we will put below $I_{CI}=1$ and
consider $A$ and $B$ as independent parameters since near $0-\pi$ transition
the ratio of these factors is not fixed.

\section{Analytical description of CPR\label{An}}

The equations (\ref{ICeq1})-(\ref{Stab}) describing the lumped junction
model can be solved analytically for some special cases.

In a vicinity of $T_{C}$ or in the limit of small thickness $d_{S}$ the
amplitude $B$ of the second harmonic is negligibly small compared to $A$,
except in a very narrow parameter range for $A=0$. As a result, we arrive at a
serial connection of two junctions with sinusoidal CPR. In this case the
net\ $I_{S}(\varphi )$ relation is given by the well-known expression
\begin{equation}
I_{S}(\varphi )=\pm \frac{A\sin (\varphi )}{\sqrt{1+A^{2}+2A\cos (\varphi )}}%
,  \label{Is1}
\end{equation}%
The shape of this dependence becomes less sinusoidal, as the magnitude of
$A$ becomes close to unity; and for $A=1$ the CPR given by Eq. (\ref{Is1})
transforms into the piecewise function
\begin{equation}
I_{S}(\varphi )=\pm \sin (\varphi /2)sign(\cos (\varphi /2)).  \label{Is2}
\end{equation}%
The minus sign in the equtions (\ref{Is1})-(\ref{Is2}) corresponds to unstable states.
In these states, the phase of the order parameter of the core s layer
differs by $\pi $ on the order parameter phase in the superconducting
electrodes. As a result, at least one of the contacts, connected in series,
would be in an unstable state.

\begin{figure}[t]
\begin{minipage}[h]{0.49\linewidth}
\center{a)\includegraphics[width=1\linewidth]{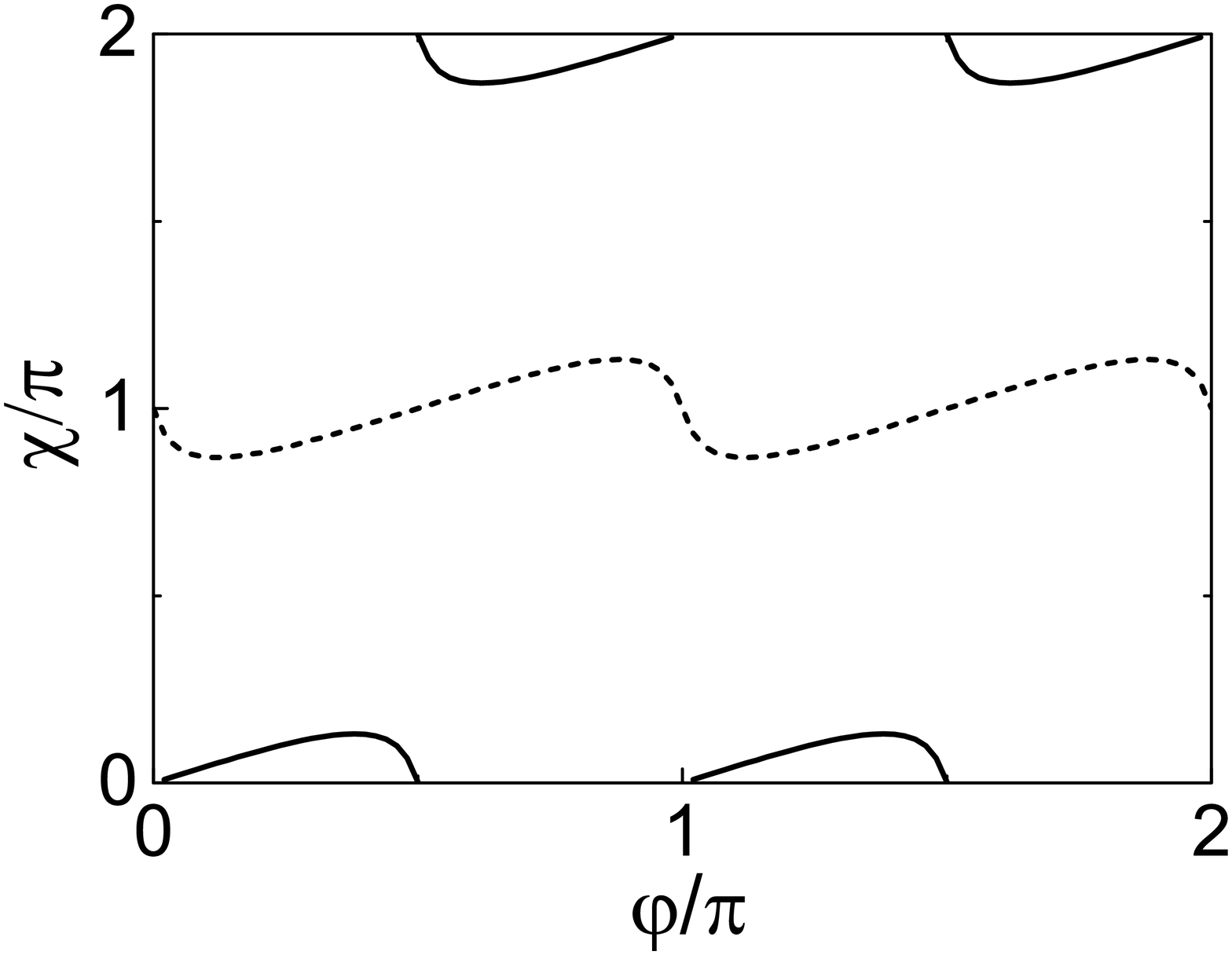}}
\end{minipage}
\hfill
\begin{minipage}[h]{0.49\linewidth}
\center{b)\includegraphics[width=1\linewidth]{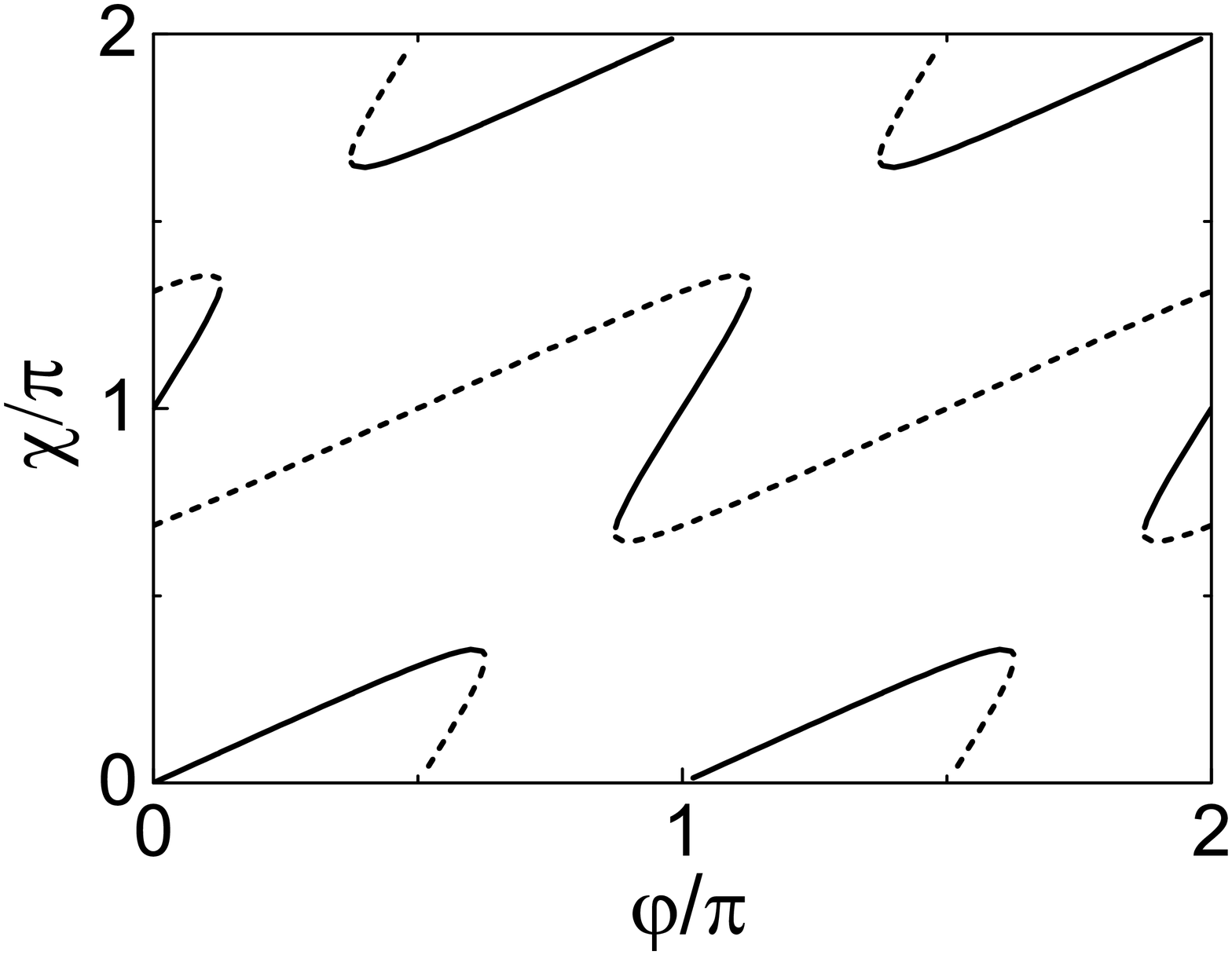}}
\end{minipage}
\par
\vfill
\begin{minipage}[h]{0.49\linewidth}
 \center{c)\includegraphics[width=1\linewidth]{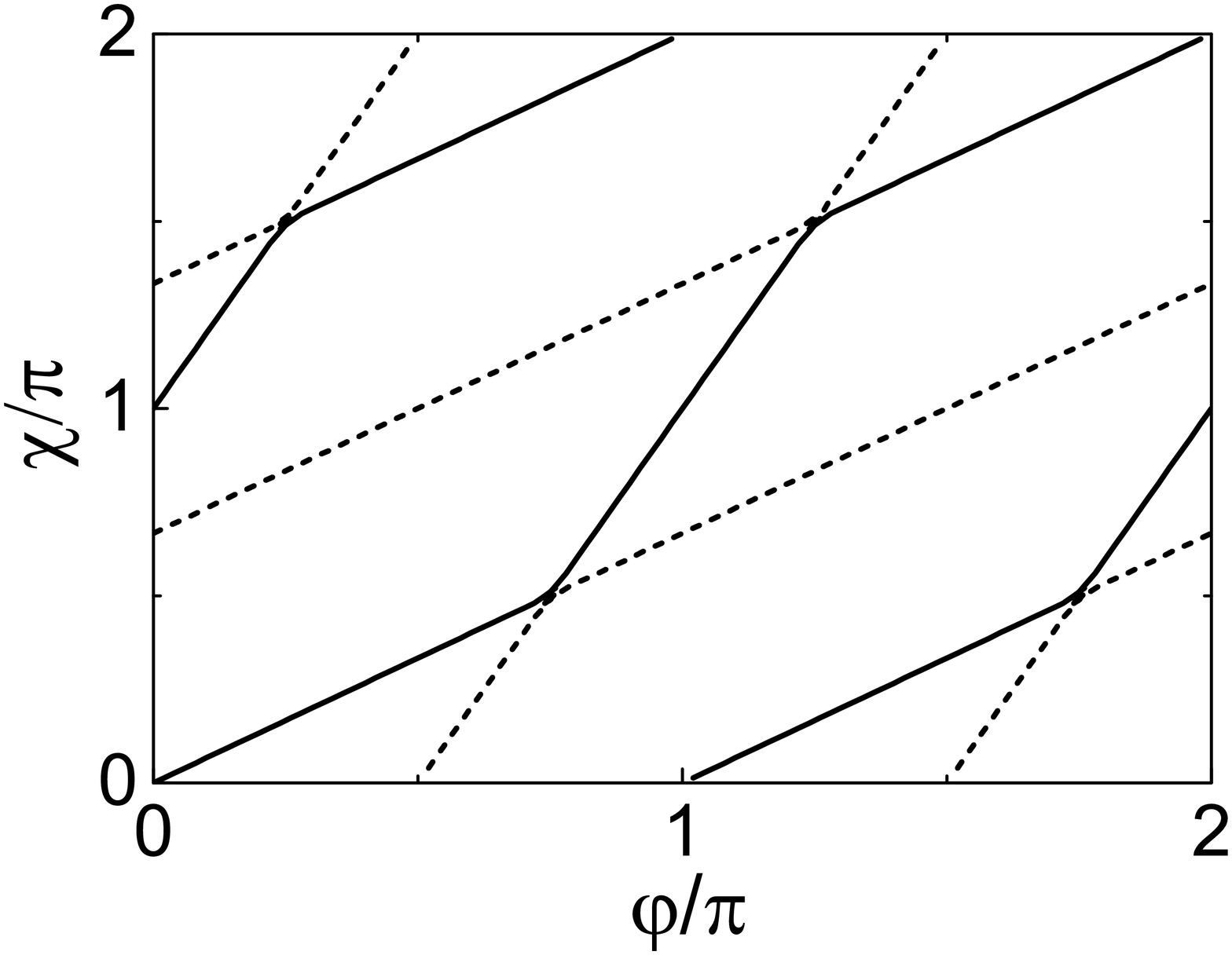}}
\end{minipage}
\hfill
\begin{minipage}[h]{0.49\linewidth}
\center{ d)\includegraphics[width=1\linewidth]{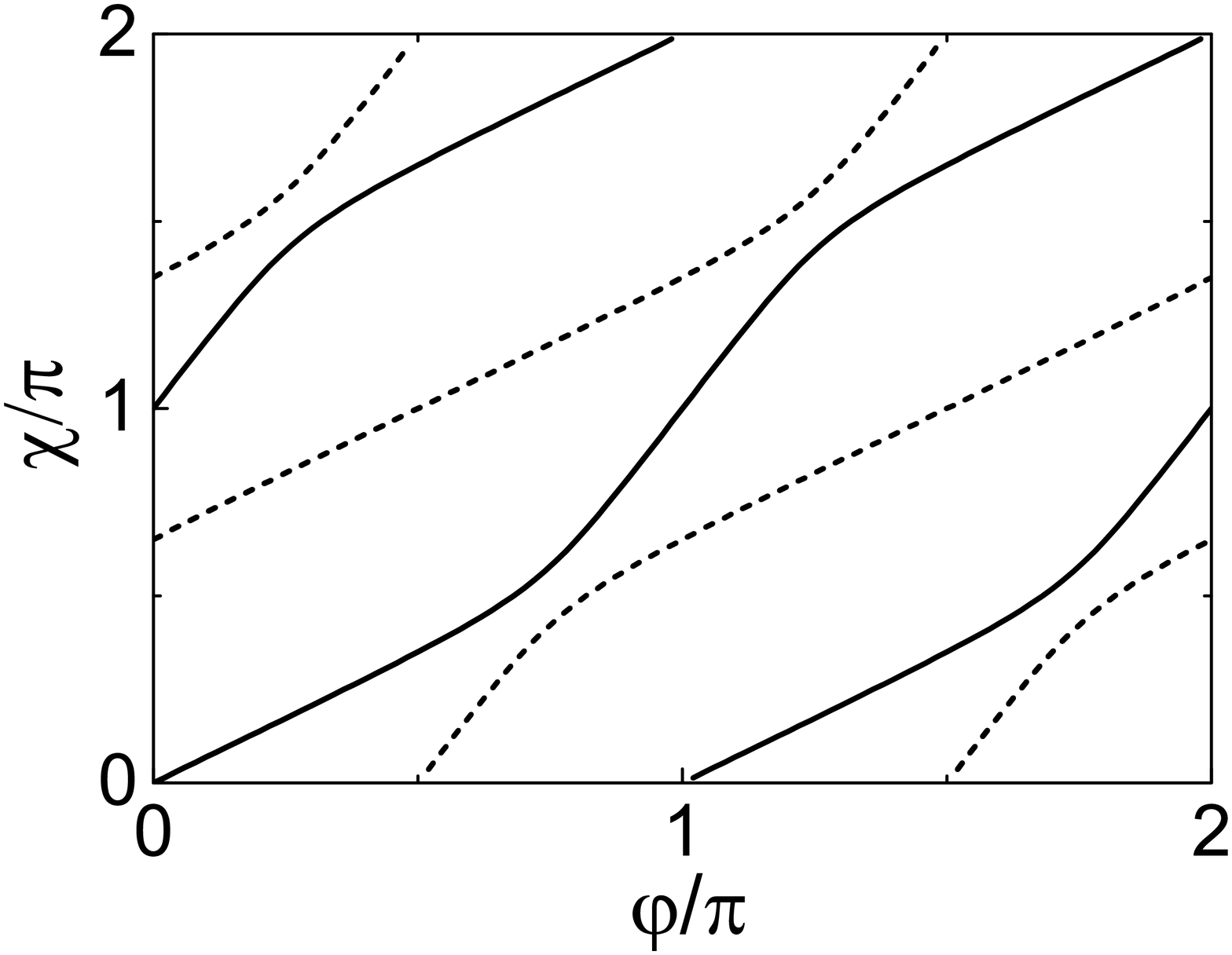}}
\end{minipage}
\caption{ The evolution of the phase $\protect\chi $ of middle s-electrode order
parameter as a function of total phase $\protect\varphi $ for the SIsFS
junction calculated in the lumped junctions model for $A=0$ and several
values of the second harmonic amplitude $B=0.4;0.9;1.0;1.1$ (panels a) - d),
respectively). The shape of $\protect\chi (\protect\varphi )$ transforms
from nonhysteretic ($B=0.4$) to hysteretic ($B=0.9$) dependence, which
occurs before merging point ($B=1.0$). The panel c) gives $\protect\chi (%
\protect\varphi )$ at merging point $B=1.0$ and the panel d) shows tunnel like
dependence $\protect\chi (\protect\varphi )$ above merging point for $B=1.1$%
. }
\label{hi_fi}
\end{figure}

At low temperatures and at large $d_{S}$ there is an interval of parameters in
the vicinity of $0$ to $\pi $ transition in which contribution to $%
I_{S}(\varphi )$ dependence from the first harmonic of sFS junction is small
compared to that from the second one. Taking $A\ll B$ in Eq. (\ref{ICeq1})
and neglecting terms proportional to $A$, we can reduce (\ref{ICeq1}) to a
fourth order equation with respect to $x=I_{S}(\varphi )=\sin (\chi )$%
\bigskip
\begin{equation}
4B^{2}x^{4}+4zx^{3}+(1-4B^{2})x^{2}-2zx+z^{2}=0,  \label{Is3}
\end{equation}%
where $z=B\sin (2\varphi )$ and phase $\chi $ is in the interval $-\pi
/2<\chi <\pi /2$ if $u=zx(z-2zx-x)>0$ and is in the range $\pi /2<\chi <3\pi
/2$ if $u<0.$ Below, we will compare the analytical expressions followed from
(\ref{Is3}) with the results of numerical solution of equations (\ref{ICeq1}%
)-(\ref{Stab}). They provide the phase $\chi $ as a function of $\varphi $
presented in Fig.\ref{hi_fi} for different values of $B$. The solid and dashed
lines in Fig.\ref{hi_fi} denote stable and unstable solutions, respectively.

In the limit $B\ll 1$ the weak place is located at the sFS part of SIsFS
structure and for $\chi (\varphi )$ one can get:

\begin{eqnarray}
\chi &\approx &\frac{B\sin (2\varphi )}{1+2B\cos (2\varphi )},~  \label{Is4}
\\
\chi &\approx &\pi -\frac{B\sin (2\varphi )}{1-2B\cos (2\varphi )}.
\label{Is4a}
\end{eqnarray}

The solution (\ref{Is4}) is stable and corresponds to the solid curves located
near $\chi =2\pi n$, as shown in Fig.\ref{hi_fi}a, calculated
numerically from (\ref{Is3}) \ for $B=0.4.$ The expression (\ref{Is4a})
gives the unstable solution shown by the dashed curve in Fig.\ref{hi_fi}a.
For $\chi \approx \pi $, the SIs tunnel junction in SIsFS device is in an
energetically unfavorable $\pi $ state, which is unstable.

Upon a further increase of the amplitude $B$ (see Fig.\ref{hi_fi}b), the
solution of equation (\ref{Is3}) becomes hysteretic in the vicinity of $%
\varphi =0+\pi n$. For $\varphi =0+\pi n$, coefficient $z=0$ and the
equation (\ref{Is3}) reduces to
\begin{equation}
\left( 4B^{2}x^{2}+1-4B^{2}\right) x^{2}=0.  \label{Is5}
\end{equation}%
and has three solutions
\begin{equation}
x_{1}=0,\quad x_{2,3}=\pm \sqrt{1-1/4B^{2}}  \label{Sol2}
\end{equation}%
For $B\leq 0.5$, only $x_{1}$ is real and $I_{S}(\varphi )$ is a
single-valued function of $\varphi .$ In the interval $B>0.5$ the $%
I_{S}(\varphi )$ dependence becomes a multi-valued function of $\varphi $
with three branches in the neighborhood of $\varphi =\pi +2\pi n.$ As will
be demonstrated below, the appearance of extra stable branch at $\chi
\approx \pi $ can be explained due to the nucleation of local minimum at $\chi =\pi $
in the $E(\varphi )$ dependence, which corresponds to energetically unfavorable stable
state of SIsFS\ structure. In this state the SIs part of the structure is in
the $\pi $-state.

With the increase of $B$, the local minimum becomes deeper and at $B=1$ the
stable branches in $I_{S}(\varphi )$ merge together (see Fig.\ref{hi_fi}%
c). At $B=1$ the critical currents of SIs and sFS parts are equal to each
other and equation (\ref{Is3}) can be simplified to
\begin{equation}
(x+z)(4x^{3}-3x+z)=0.  \label{Is6}
\end{equation}%
Taking into account the restrictions on the intervals $\chi $ in equation (%
\ref{Is3}), we can write the solutions of (\ref{Is6}) in the form

\begin{eqnarray}
\chi &=&\pi +2\varphi,  \label{Sol3} \\
\chi &=&2/3\varphi +2/3\pi n,  \label{Sol4}
\end{eqnarray}%
where $n=0,1,2.$

The equalities (\ref{Sol3}), (\ref{Sol4}) provide a set of four intersecting
linear relationships between phases $\chi \ $\ and $\varphi.$ The
intersection points of these lines, as shown in Fig.\ref{hi_fi}c, divide each of them
into stable and unstable regions. Deviation of $B$ from unity in the
direction of smaller values leads to a separation of this linear solutions, as
shown in Fig.\ref{hi_fi}b. Contrary to this, with the increase of $B$, the weak place
is shifted towards SIs tunnel junction resulting in growth of $\chi \ $\ with $%
\varphi $ (see Fig.\ref{hi_fi}d). The larger the value of $B$, the smaller is
deviation from linear relation $\chi =\varphi +\pi n$ between $\chi \ $\ and
$\varphi .$ It is important to note that the SIs tunnel junction can be both in a $%
0$-state $(n=0)$ and $\pi $-state $(n=1)$ depending on the condition in
which there is the sFS contact.


\section{\protect\bigskip Numerical description of CPR\label{Num}}

\begin{figure}[t]
\begin{minipage}[h]{0.49\linewidth}
\center{a)\includegraphics[width=1\linewidth]{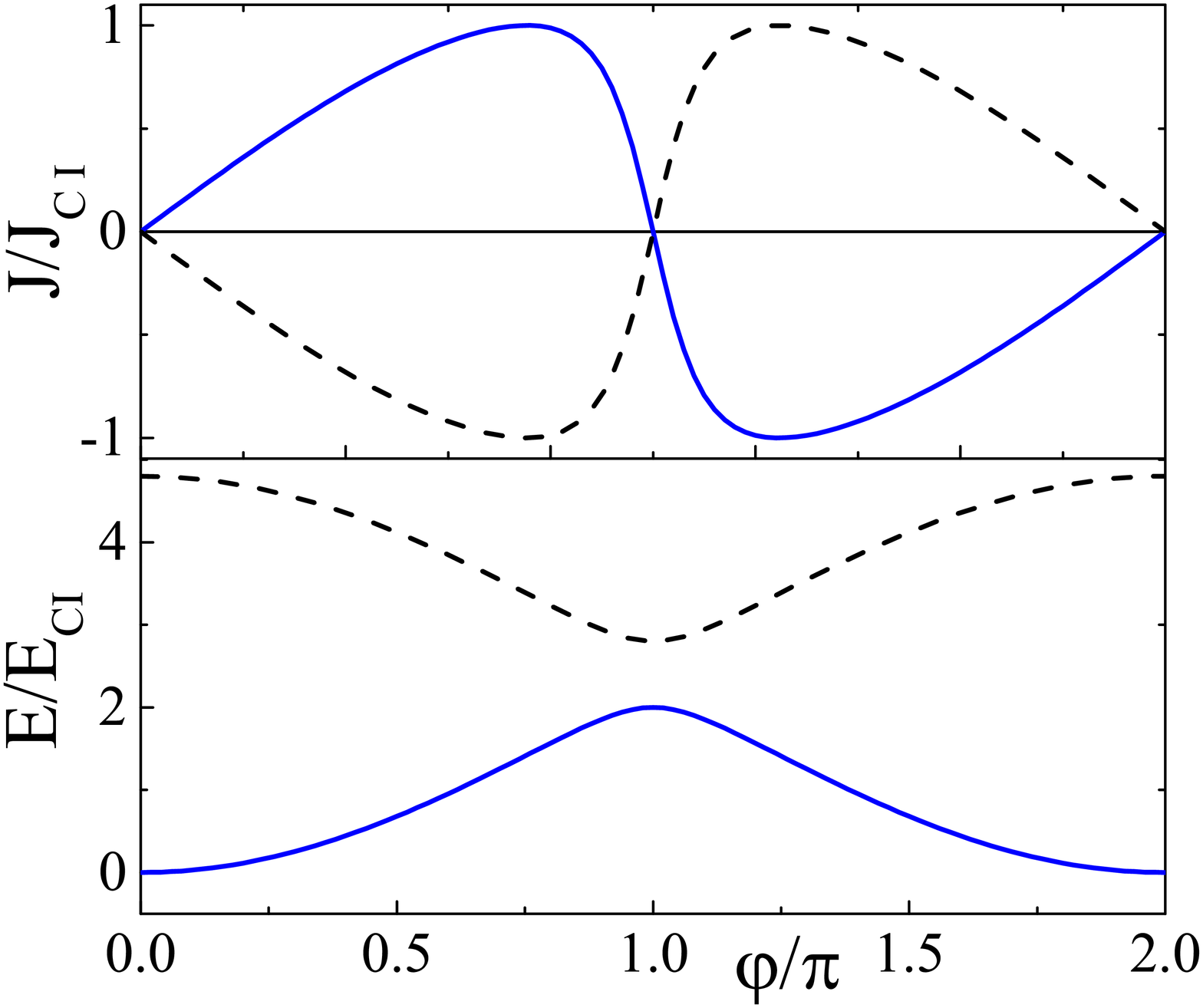}}
\vspace{-6 mm}
\end{minipage}
\hfill
\begin{minipage}[h]{0.49\linewidth}
\center{b)\includegraphics[width=1\linewidth]{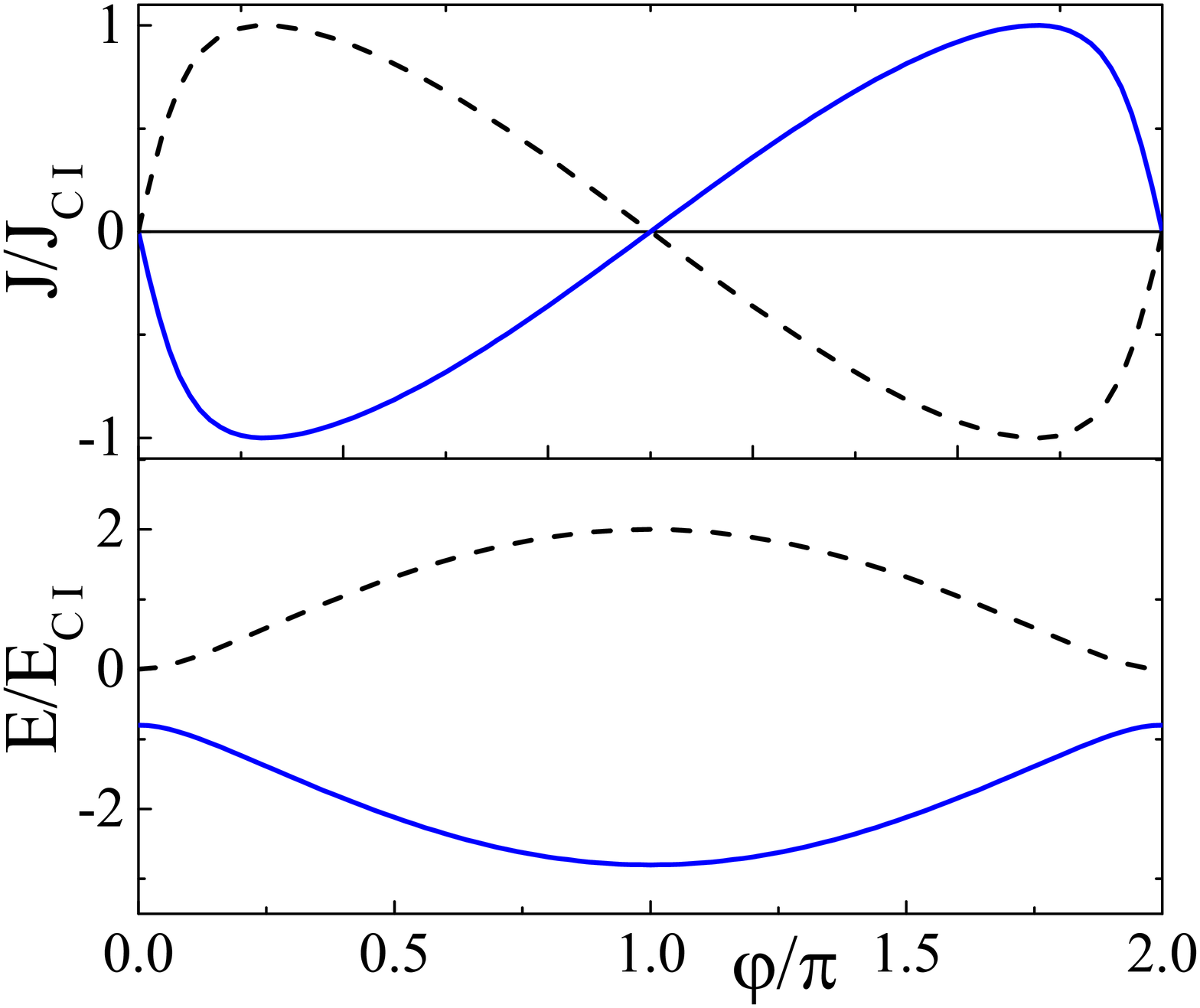}}
\vspace{-6 mm}
\end{minipage}
\vfill
\begin{minipage}[h]{0.49\linewidth}
\center{c)\includegraphics[width=1\linewidth]{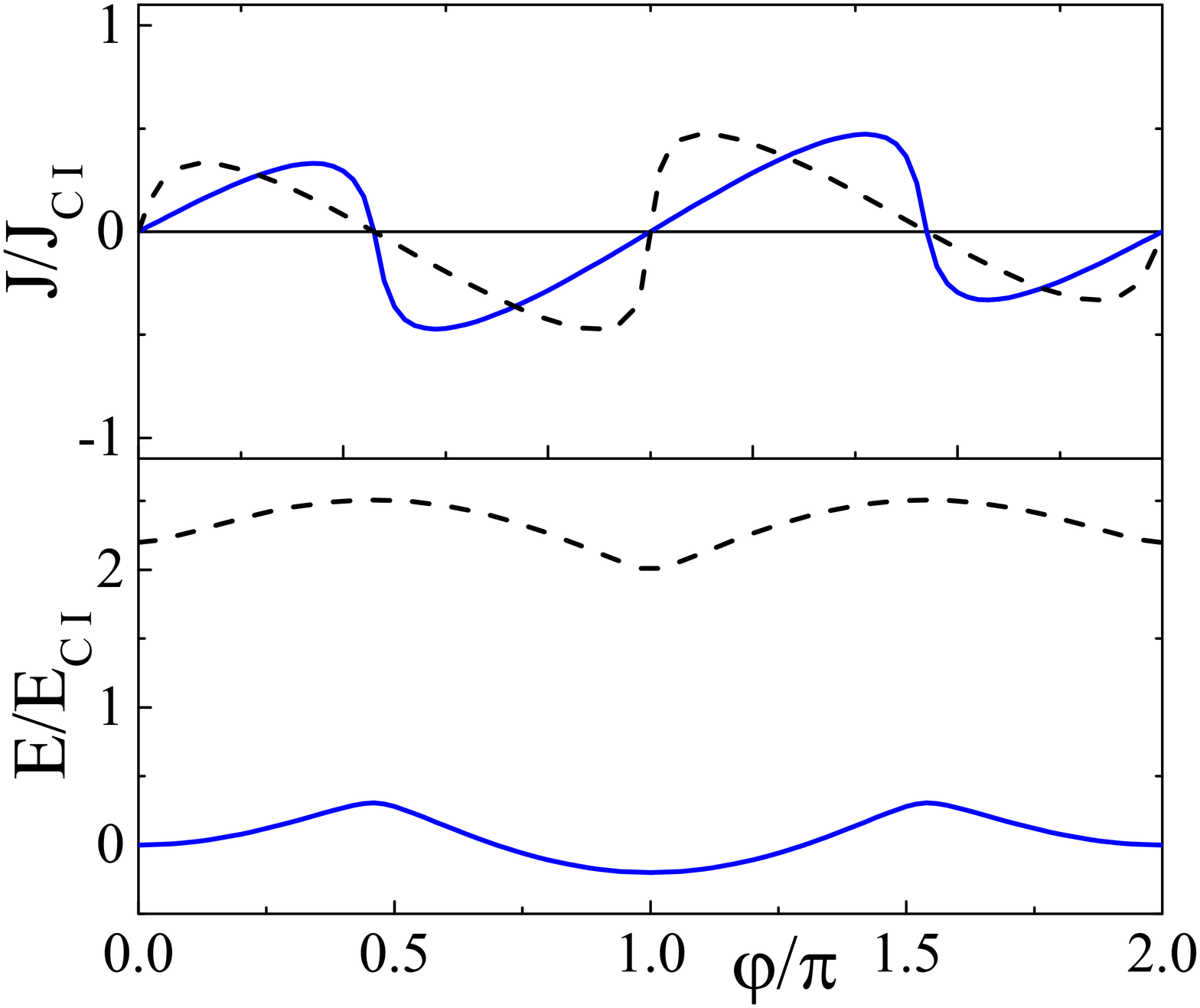}}
\vspace{-2 mm}
\end{minipage}
\hfill
\begin{minipage}[h]{0.49\linewidth}
\center{d)\includegraphics[width=1\linewidth]{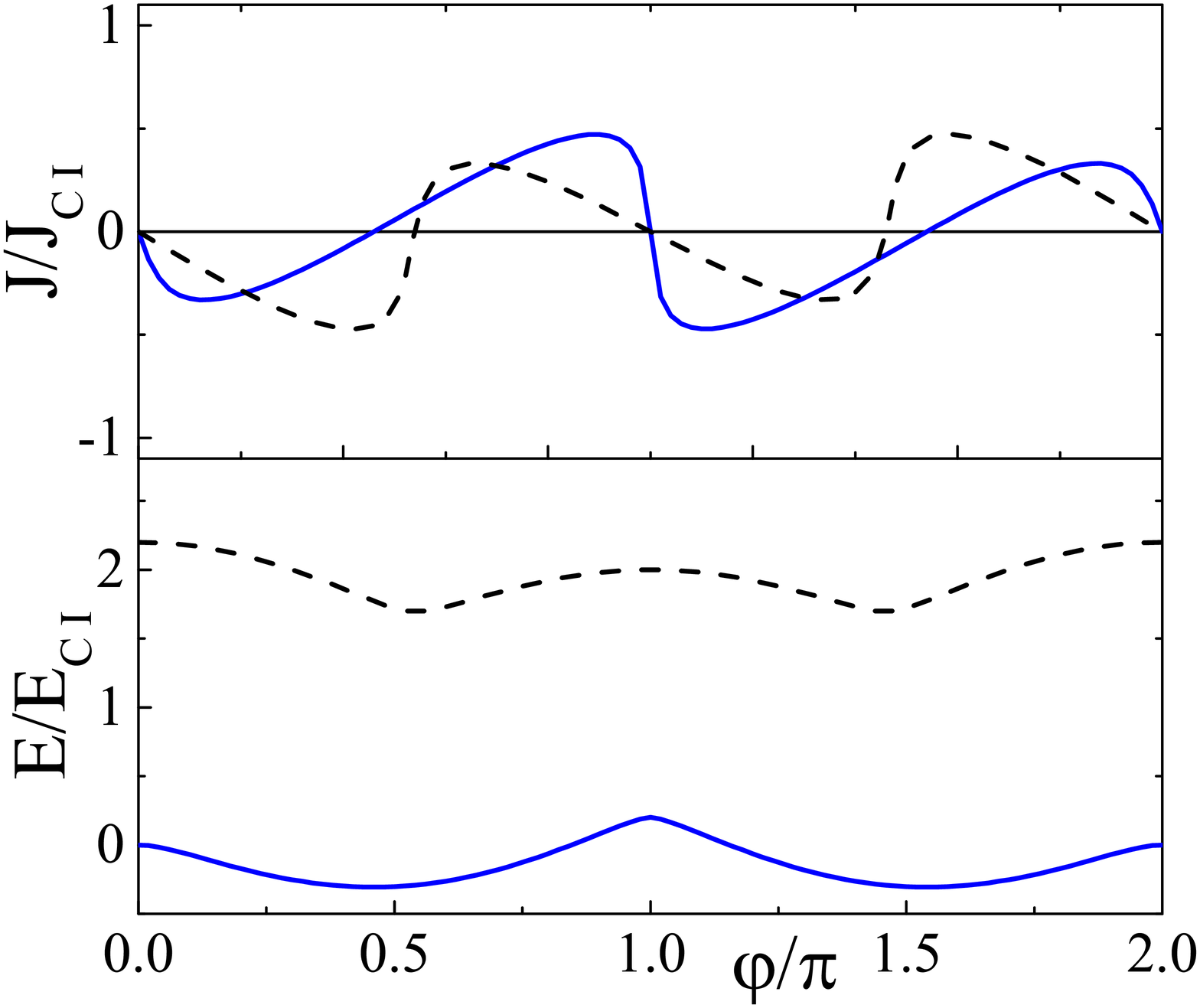}}
\vspace{-2 mm}
\end{minipage}
\caption{(Color Online) The current-phase (top panel) and the energy-phase (bottom
panel) relations of SIsFS junction calculated in the lumped junctions model
for combinations of amplitudes $A$ and $B$ provided trivial
single-valued shape of CPR: a) $0$-ground state at $A=1.4$, $B=0$, b) $%
\protect\pi $-state at $A=-1.4$, $B=0$, c) $0$-$\protect\pi $-ground state
at $A=0.1$, $B=0.4$, d) $\protect\varphi $-ground state at $A=0.1$, $B=-0.4$%
. It is seen that, serial connection of SIs and sFS junctions provides
significant deviations of CPR from sinusoidal shape even in the absence of
the second harmonic.}
\label{CPR_conv}
\end{figure}
\begin{figure}[t]
\center{\includegraphics[width=0.85\linewidth]{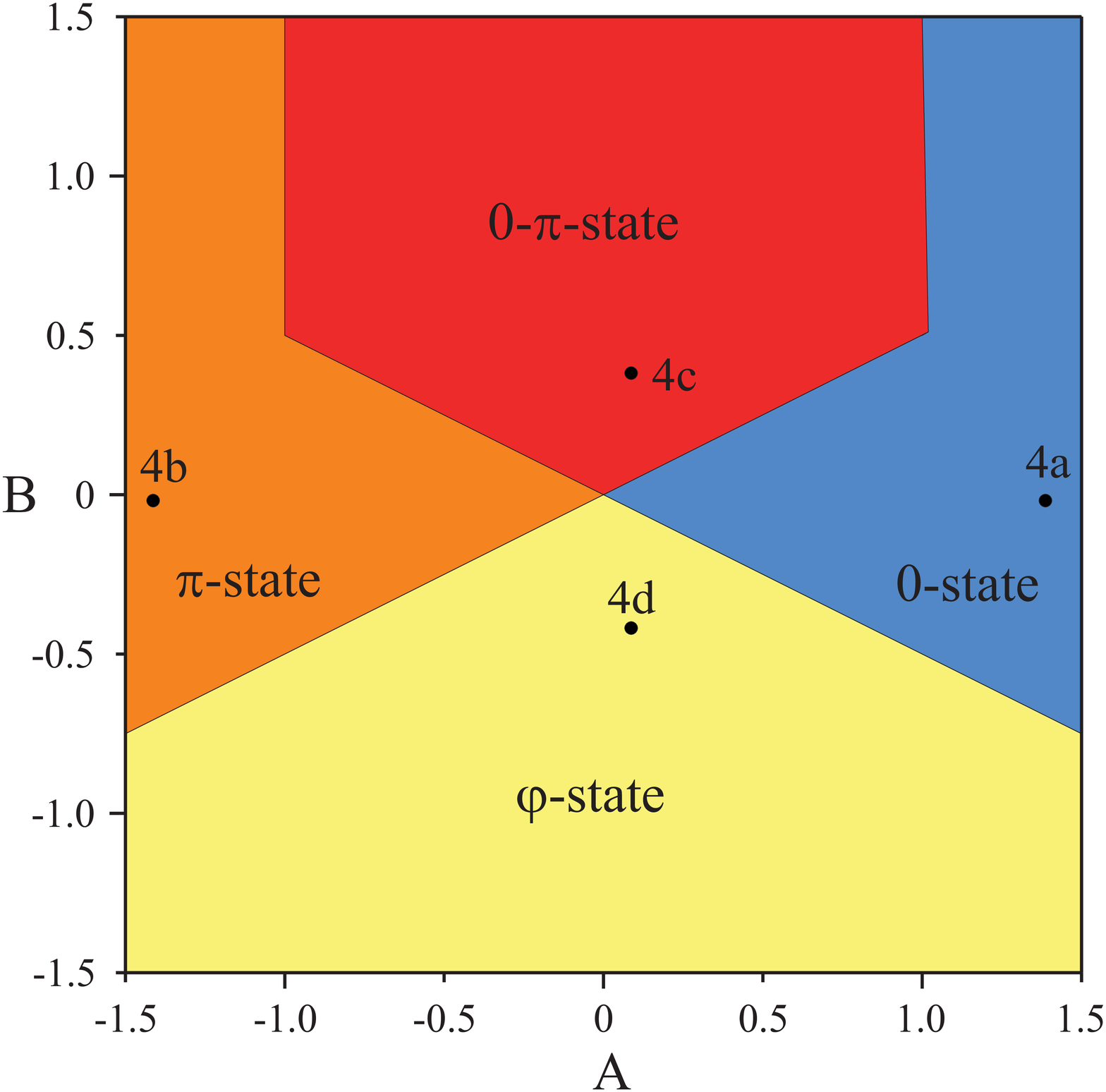}}
\caption{(Color Online) Ground state distribution in the SIsFS structures on
(A, B) phase plane. }
\label{PhaseDiagram1}
\end{figure}
For the arbitrary values of amplitudes $A\ $and $B$ we solve equations (\ref%
{ICeq1})-(\ref{Stab}) numerically. We study the possible transformations of
CPR shapes in the area $\left\vert A\right\vert \leq 1.5,$ $\left\vert
B\right\vert \leq 1.5,$ that encloses the values of $A\ $and $B$ at which
the transition occurs between the $0$-and $\pi $-states in SIsFS junction. A
finite amplitude $A$ results in an increase of the number of possible shapes
of CPR in comparison with the discussion made above.

Below we give a classification of the physical states implemented in the
SIsFS structures based on two criteria. The first one includes
the classification of the ground states. Those are the states that meet the
requirements of the minimum of the functional $E(\varphi )$ and of the
stability (Eq.\ref{Stab}). In addition we don't take into account the minima,
which has energy larger than the energy of any other branch at the same $%
\varphi $. The second criterion specifies information about the shape of $%
I_{S}(\varphi )$ curves. It gives the number, $k,$ of stable branches of $%
I_{S}(\varphi )$ unconnected with one another, as well as the number, $m,$
of possible jumps caused by the transition between these branches arising
during $\varphi $ swiping in the interval $0\leq \varphi \leq 2\pi .$

There are four possible types of ground states of SIsFS structure. Their
examples, as well as the corresponding $I_{S}(\varphi )$ relationships are
shown in Fig. \ref{CPR_conv}. The implementation of $0$ or $\pi $ states
depends mainly on the relationship between the amplitudes $2|B|$ and $|A|$
(see Fig.\ref{PhaseDiagram1}). For $|A|>2|B|$ the SIsFS structure has the
single ground state at $\varphi =0$ at positive $A,$ and at $\varphi =\pi $
at negative $A$. Figures \ref{CPR_conv}a and \ref{CPR_conv}b reveal the $%
E(\varphi )$ and $I_{S}(\varphi )$ calculated for $B=0$ and $A=\pm 1.4,$
respectively. At $A=1.4$, the minimum of $E(\varphi )$ is achieved at $\varphi
=0+2\pi n,$ while for $A=-1.4$ it is shifted towards $\varphi =\pi +2\pi n.$ We
have identified these ground states as $0$ and $\pi $, respectively. These
types of CPRs can be also observed in the regular SFS and S-F/N-S junctions
\cite{RevG}. However, in the SIsFS junctions the shapes of $I_{S}(\varphi )$
significantly deviate from sinusoidal one even in the absence of the second
harmonics $(B=0)$.

Figures \ref{CPR_conv}c and \ref{CPR_conv}d show the $E(\varphi )$ and $%
I_{S}(\varphi )$ calculated for $B=\pm 0.4$ and $A=0.1$ respectively. For
positive $B$ the $E(\varphi )$ curve (see Fig.\ref{CPR_conv}c) has two minima at
$\varphi =0$ and $\varphi =\pi .$ We classify this situation as $0 $-$\pi $
ground state \cite{Radovic1, Radovic2}. The diagram Fig.\ref{PhaseDiagram1} shows that a $0$-$\pi $
ground state exists if $|A|<2|B|$ and $|A|<1.$ The case $|A|>1$ is less
trivial and will be discussed below.


For $|A|<2|B|$ and $B<0$, the $E(\varphi )$ curve (see Fig.\ref%
{PhaseDiagram1}) reaches a global minimum at some arbitrary phase $\varphi
=\pm \varphi _{g},$ which does not coincide with both $\varphi =0$ and $%
\varphi =\pi .$ Figure \ref{CPR_conv}d demonstrates an example of this
situation realised for $A=0.1$ and $B=-0.4$. For small $|B|\ll 1$,
the properties of SIsFS junction are similar to that of so-called $\varphi $%
-junction \cite{Buzdin}, such that the magnitude of $\varphi _{g}$ can be any
value in the range $[0,\pi ]$. With increase of $|B|$, the interval available
for $\varphi _{g}$ diminishes and for $|B|$ $\gg 1$ it asymptotically
converges to $\pm \pi /2.$ $\,$It is necessary to note that the condition $B<0$
can be realized in junctions with a complex internal structure of their weak
link region \cite{Buzdin, Pugach1, Gold, Bakurskiy3, Gold2, GoldButterfly}.

Contrary to the result presented in Fig.\ref{compare}, all the $%
I_{S}(\varphi )$ dependencies shown in Fig. \ref{CPR_conv} are single-valued
functions of $\varphi .$ These types of current-phase relations exist only in
the limited area in the $(A,B)$ phase plane. It means that the phase diagram Fig.%
\ref{PhaseDiagram1} is rather crude. It requires a further clarification of
the boundaries separating the areas of single-valued and multi-valued
current phase relations.

For further determination of possible CPR, we need to introduce additional parameters.
They are indices $k$ and $m$. As defined above, the index $k$
counts the number of stable branches of $I_{S}(\varphi )$ including ground
states and unconnected with another branch geometrically, so that\ switching
of the system from one branch to another is possible only through a phase
slip. The index $m$ gives the number of possible jumps caused by the
transition between these branches arising during $\varphi $ increase in the
interval $0\leq \varphi \leq 2\pi .$ We determine it as the number of phase
slips during continuous increase of phase $\varphi $, starting from the
position at the ground state. The counting ends at a value of $\varphi $ that is
different from the initial one by $2\pi $ even if the systems stays on the
other branch with further increase of $\varphi $. In this way, the number $%
m_i$ is found for each ground state. The resulting index $m = \sum{m_i}$ is
a sum over existing ground states.

\begin{figure*}[t]
\center{\includegraphics[width=0.8\linewidth]{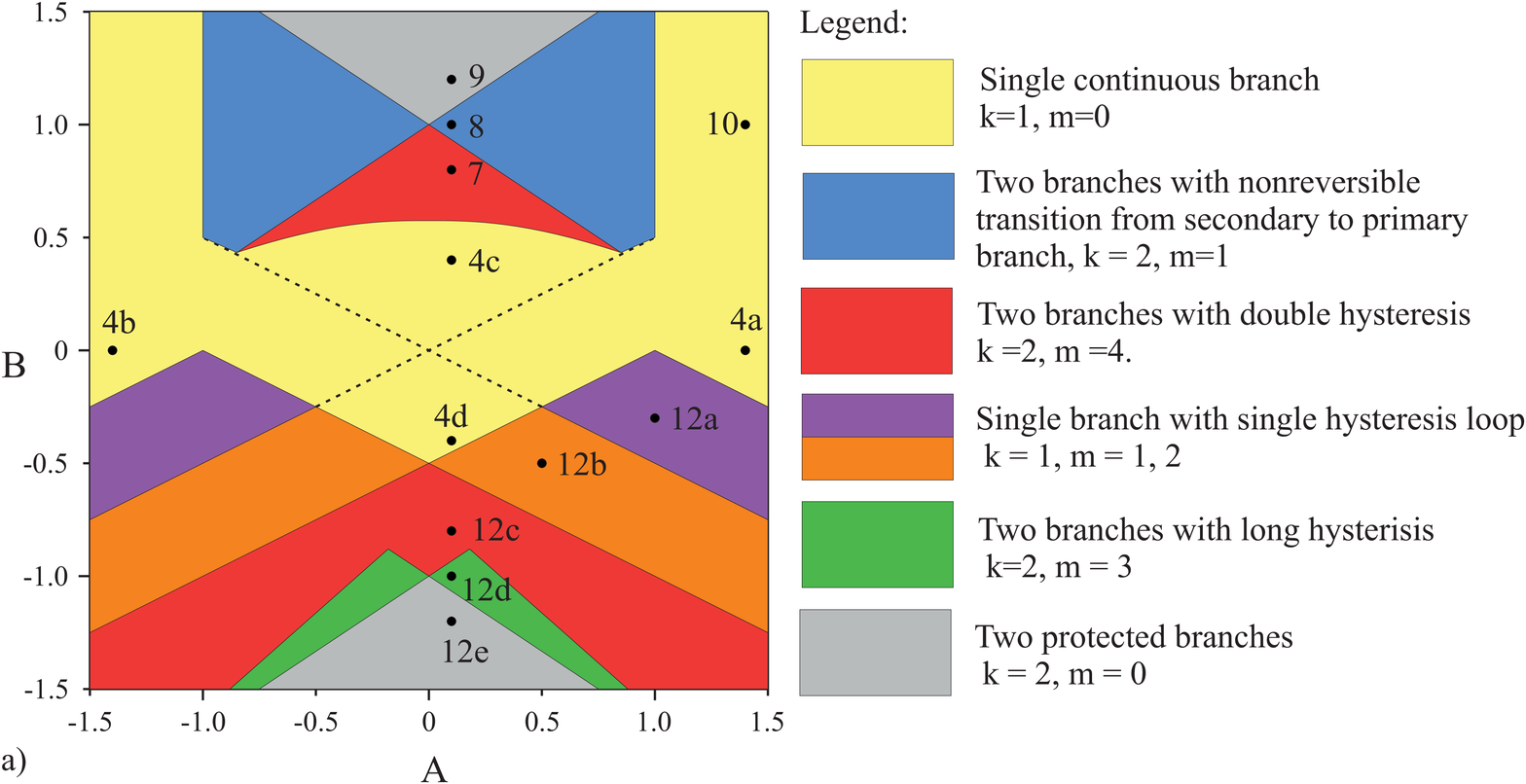}}
\caption{(Color Online) Distribution of indices of SIsFS
junction CPR in the (A, B) phase plane. The legend reveals correspondence between
color and the indices: $k$ is the number of stable branches of $%
I_{S}(\protect\varphi )$ including ground states, which and unconnected with
each other geometrically; $m$ gives the number of possible jumps caused by
the transition between these branches arising during $\protect\varphi $
increase in the interval $0\leq \protect\varphi \leq 2\protect\pi .$ The dashed
lines define the boundary between different types of ground states. The
filled black circles in the plane are the points, at which the CPR presented
in Fig.4 and Fig.7-Fig.10 have been calculated in the frame of lumped
junctions model. The number of corresponding figure is written near the
circles. The shapes of CPR for negative $B$ are presented in Appendix in
Fig. 12. }
\label{PhaseDiagram2}
\end{figure*}

\begin{figure}[h]
\center{\includegraphics[width=0.9\linewidth]{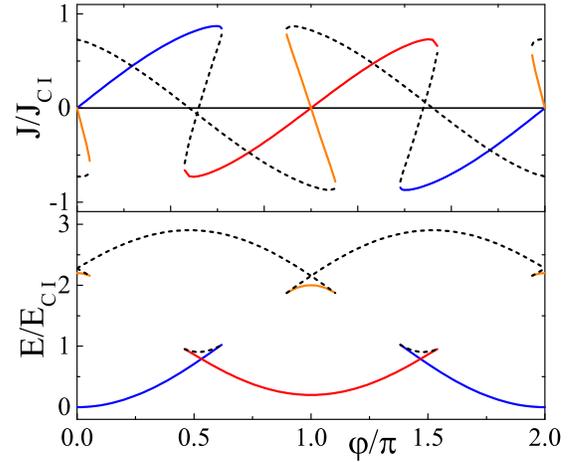}}
\caption{ (Color Online) The current-phase relation (top panel) and the energy-phase
relation (bottom panel) for the SIsFS structure in hysteretic state $k=2$, $%
m=4$ calculated for $A=0.1$ and $B=0.8$. The solid lines correspond to stable
states. The blue line is a branch including ground state $\protect\varphi=0$,
the red line corresponds to a ground state $\protect\varphi=\protect\pi$, and two
orange lines show energetically higher states with $\protect\pi$ shift
across the SIs tunnel junction. The dashed lines show unstable states.}
\label{State1}
\end{figure}

\begin{figure}[h]
\center{\includegraphics[width=0.9\linewidth]{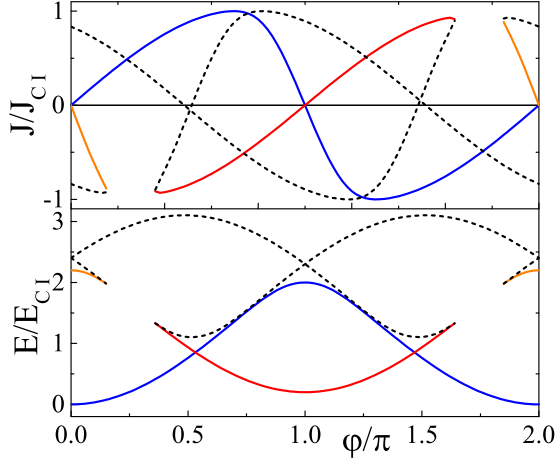}}
\caption{ (Color Online) The current-phase relation (top panel) and energy-phase
relation (bottom panel) for the SIsFS structure in the state with primary
branch $k=2$, $m=1$ calculated for $A=0.1$ and $B=1.0$. The solid lines
correspond to stable states. The blue line corresponds to a ground state at $%
\protect\varphi=0.$ It is the primary branch, which is stable in the whole range of variation $0\leq \protect\varphi \leq 2 \protect\pi%
. $ The red line is the secondary branch, stable parts of which exist only in some
interval of $\protect\varphi$ in a vicinity of $\protect\varphi=\protect\pi$%
. The orange line located nearby $\protect\varphi=0$ shows energetically higher
states with the $\protect\pi$ shift across the SIs tunnel junction. The dashed lines
show unstable states.}
\label{State2}
\end{figure}

\begin{figure}[h]
\center{\includegraphics[width=0.9\linewidth]{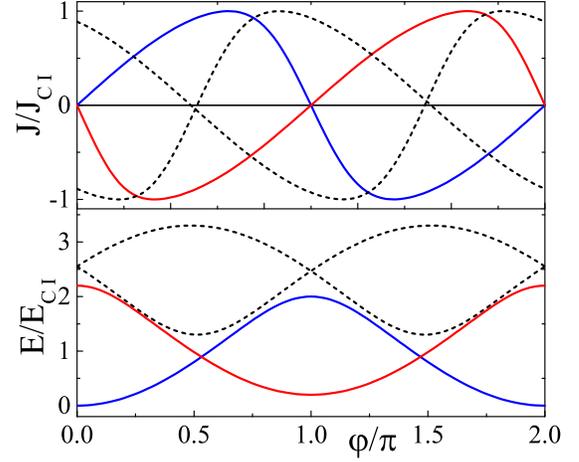}}
\caption{ (Color Online) The current-phase relation (top panel) and the energy-phase
relation (bottom panel) for the SIsFS structure in the state with two
independent protected branches $k=2$, $m=0$ calculated in the lumped
junctions model for $A=0.1$ and $B=1.2$. Solid lines show the stable states.
Blue and red lines correspond to the branches having ground state at $%
\protect\varphi=0$ and $\protect\varphi=\protect\pi,$ respectively. Both
branches exist for all $\protect\varphi$ in the range $0\leq \protect\varphi %
\leq 2 \protect\pi.$ By an adiabatic change of the phase $\protect\varphi$
it is impossible to switch between these two branches. The dashed lines show
unstable states.}
\label{State3}
\end{figure}

\begin{figure}[h]
\center{\includegraphics[width=0.9\linewidth]{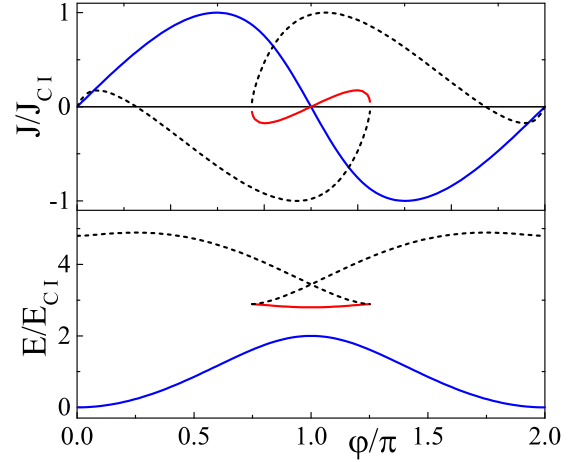}}
\caption{ (Color Online) The blue lines show current-phase relation (top
panel) and the energy-phase relation (bottom panel) for the SIsFS structure with
 multivalued CPR shape ($k=1$, $m=0$) having single ground state at $%
\protect\varphi=0$. The red lines show the analogous curves for
energetically unfavorable state. It is seen that a metastable state at $%
\protect\varphi=\protect\pi$ also is possible. The dashed lines show unstable states.
Calculation is done in the lumped junctions model for $A=1.4$ and $B=1.0$.}
\label{State4}
\end{figure}

The classification is summarized in Fig. \ref{PhaseDiagram2}, which presents the
information on the number of hysteretic regions in $I_{S}(\varphi )$
dependence, and also on the mutual positions of ground states and phase jumps.
The filled black circles in the plane are the points, at which the CPR
presented in Fig.\ref{CPR_conv} and Figs. \ref{State1}-\ref{State4} have been calculated in the frame of
lumped junctions model. The number of corresponding figure is written near
the circles.

Figs. \ref{State1}-\ref{State4} demonstrate the main classes of
the current-phase relations. In the diagram presented in Fig. \ref{PhaseDiagram2} they
are marked by different colors. Each panel in Figs.\ref{State1}-\ref{State4}
gives $I_{S}(\varphi ) $ and $E_{S}(\varphi )$ dependencies calculated
numerically from equation (\ref{ICeq1}). As in Fig.\ref{hi_fi}, the dashed
black lines show unstable states. Different colors of solid lines correspond
to the different branches of stable solutions. From Fig. \ref{PhaseDiagram2} it
is seen that for positive $B$, $I_{S}(\varphi )$ transforms into a
multi-valued function for $B\gtrsim 0.5$ and $|A|\lesssim 0.75$.

Typical $I_{S}(\varphi )$ and $E_{S}(\varphi )$ curves for the area $7$ in
Fig. \ref{PhaseDiagram2}a are shown in Fig.\ref{State1}. They have been
calculated for $A=0.1$ and $B=0.8.$ The current-phase relation consists of
two stable branches leading to $k=2$. In the domain $0\leq \varphi \leq 2\pi
$ phase sweep from $0$ to $2\pi $ 
must lead to two hops between stable branches for each of the two
available ground states resulting in $m=4.$
It is necessary to note that in this area of parameters $A$ and $B$ there
are some additional stable branches at higher energies (orange lines on Fig.%
\ref{State1}, which don't correspond to the ground state. Due to large energy
difference between these states it is impossible to switch between them by
an adiabatic change of the phase $\varphi $. 

Figure \ref{State2} gives an example of $I_{S}(\varphi )$ and $E_{S}(\varphi
)$ curves typical for area $8$ in $A-$ $B$ plane in Fig. \ref{PhaseDiagram2}.
They have been calculated for $A=0.1$ and $B=1$. At $B=1$ the shapes of $%
I_{S}(\varphi )$ and $E_{S}(\varphi )$ dependencies exhibit a
transition to a state with $k=2$ and $m=1$. With increase of $B$, the stable
branches corresponding to the minimum energy (marked by blue in Fig. \ref%
{State1}) tend to connect to the stable branches corresponding to the maximum
energy $($marked by orange in Fig. \ref{State1} in a vicinity of $\varphi
=\pi ).$ For particular case of $A=0.1$ shown in Fig. \ref{State2}, the
connection has completed at $B=1$ resulting in formation of the continuous $%
I_{S}(\varphi )$ and $E_{S}(\varphi )$ dependencies without any hysteresis. For
finite $A$, the minima of $E(\varphi )$ at $\varphi =0$ and $\varphi =\pi $
have different depth $(E(0)<E(\pi ))$ and corresponding merging points split
at $B_{1,2}=1\pm 2/3$ $A$. In the interval $B_{1}\leq B\leq B_{2}$ and at $%
|A|<1$, the branch of $E(\varphi )$ passing through a deeper minimum is
already continuous at every $\varphi $, while the second branch of $%
E(\varphi )$ exists only in some intervals of $\varphi $ (See Fig. \ref%
{State2}). An escape of phase $\varphi $ from this intervals leads to the jump
of $E$ on a more stable $E(\varphi )$ branch. After that the SIsFS junction
can't be adiabatically switched back into the previous state.

The next transition to the state with $k=2$ and $m=0$ occurs for
$|A|<1$ and the amplitude $B$ exceeds $B_{2}$ (Fig. \ref{State3}). In this area
of amplitudes the weak place is located at SIs junction, while SIsFS structure
can stay either in $0$- or in $\pi $-state. One of the two energetically
favoured states corresponds to the global minimum of the energy at $\varphi
=0$, while the second corresponds to a local minimum at $\varphi =\pi $. The
magnitudes of $E(\varphi )$ at $\varphi =0$ and $\varphi =\pi $ are slightly
different. These states are protected from each other in the
sense that a transition from one of them to another is not possible with a
continuous adiabatic phase change of  $\varphi $. To switch SIsFS junction
between the $0$- and $\pi $-states, one should increase a bias current across
the junction to a value larger than the critical current of sFS part of the
structure.

Finally, the region with $\left\vert A\right\vert>1$ corresponds to the
dependence shown on Fig. \ref{State4}. The CPR in this state also has two
branches with minima on $E(\varphi)$ dependence. However, the split between
branches is too large, and local minimum of upper branch is energetically
higher than maximum of lower branch. Thus, the upper local minimum can not
be declared as a possible ground state, and we don't count this branch in
indices $k$ and $m$. In this way, we call the states at $\left\vert
A\right\vert>1$ as $0$- and $\pi$- states on phase diagram Fig. \ref%
{PhaseDiagram1} and consider it as the state with single branch $k$=1 on
Fig. \ref{PhaseDiagram2}a.

\begin{figure}[t]
\center{\includegraphics[width=0.90\linewidth]{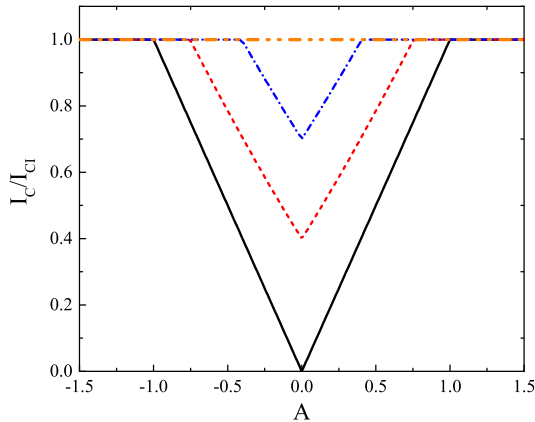}} \vspace{1 mm}
\caption{(Color Online) The dependence of critical current $I_C$ on amplitude of
the first CPR harmonic $A$ during $0$-$\protect\pi$ transition in SIsFS
junction. The $0$-$\protect\pi$ transition is defined as a change of $A$
sign under the condition of fixed second harmonic amplitude $B$. Solid
black, dashed red, dash-dot blue and dash-dot-dot orange lines correspond to
$B=0; 0.4; 0.7$ and $B\geq1$, respectively. }
\label{CritCurr}
\end{figure}

The current-phase relations with negative sign of the second harmonic amplitude $B$ are less common and require the realization of complex F-region
consisting from a number of layers. \cite{Buzdin, Pugach1, Gold, Bakurskiy3, Gold2}.
Therefore, we shift the discussion of the classification of the states shown at the bottom half of the diagram in Fig. \ref{PhaseDiagram2} to Appendix A.

The above classification of CPR may help to interpret the
experimental data in the SIsFS structures near $0$- to $\pi $-transitions. In standard SFS junctions, $0$-$\pi$ transitions manifest themselves as dips in the  $I_C(d_F)$ or $I_C(T)$ dependencies. Experimental results for SIsFS junctions demonstrate similar behaviour in the regime of small thickness $d_s$. However, such  dips disappear for large $d_s$ (see Ref. \cite{Ruppelt}).

To explain this effect, we consider the dependence of the critical current $I_C$ on the first harmonic amplitude $A$ for several fixed values of the second harmonic $B$ as shown in Fig. \ref{CritCurr}. In the absence of the second harmonic (the solid line), the pronounced dip of $I_C$ is visible indicating $0$- to $\pi $-transition. In the parameter range within the dip, the weak link is shifted from the tunnel barrier I to the ferromagnetic layer F. Far from the $0-\pi$ transition the magnitude of $I_C$ is independent on $A$ and equals to the critical current of the SIs tunnel junction, where the weak link is located.

With the increase of $B$ the CPR deforms and additional branches start to appear. As a consequence, the dips at $I_C(A)$ curves gradually decrease (see the red dashed and the blue dash-dot lines on Fig. \ref{PhaseDiagram2}). Finally, at $B>1$ the dip vanishes and the weak link is always located at the tunnel barrier. As a result, $I_C$ remains constant across the $0-\pi$ transition (the orange dash-dot-dot line).

As follows from the above discussion, the standard approach for detection of $0$-$\pi $
transitions, based on measurements of $I_{C}R_{N}(d_{F})$ dependencies,
breaks down in SIsFS junctions at low temperatures and large s-layer thickness $\gtrsim 3 \xi_{S}$.
Detection of such transitions requires phase-sensitive experiments \cite{Frolov}.

\section{Conclusion}

As follows from our analysis, the CPR in the SIsFS structures
is qualitatively different from that in regular SFS junctions. We
have demonstrated that the classification of the various CPR types requires the use of two
indices. One of them, $k,$ indicates the number of the existing ground
states, while the other, $m,$ defines the number of current leaps occurring
during variation of the phase difference $\varphi $ in each of these ground states
from $0$ to $2\pi $. We have also shown that the values of these
indices depend on the ratio between the amplitudes of the first, $A,$
and second, $B,$ harmonics in CPR\ of sFS part of SIsFS\ junction. We have
identified the areas in the $A$-$B$ plane corresponding to all possible
combinations of pairs of these indices, as well as the typical shapes of the
CPR for each of these areas. We have shown that some of the found states are protected.
The example is given in Fig. \ref{State3}, which depicts two CPR in the protected state
with indices $k=2$ and $m=0.$ In this case the SIsFS structure can stay
either in $0$- or in $\pi $-ground state, with only slight difference between the magnitudes of $E(\varphi
)$ at $\varphi =0$ and $\varphi =\pi $. Furthermore, a transition from one ground state to another is not possible by
a continuous adiabatic variation of the phase $\varphi$. Our preliminary analysis
done in the frame of RSJ model confirms that this property is conserved even
in a dynamic regime, despite there is a voltage drop across the SIsFS
junction and both $\chi $ and $\varphi $ are time dependent. To switch SIsFS
junction between the $0$- and $\pi $-states, one should increase a bias
current across the junction to a value larger than the critical current of the
sFS part of the structure. More detailed consideration of switching between
 protected CPR branches will be done elsewhere.
 
Note that there is some similarity between the considered
properties and the effects found in the topological systems based on
multi-terminal Josephson junctions \cite{Giazotto, LinderTop, GiazottoTh}.
In the latter case, different topological states correspond to different
distributions of phase differences between the terminals. In
SIsFS junctions intermediate electrode can be considered as additional
terminal embedded into the SIFS weak link. The resulting states
of SIsFS contacts become separated and any transition between them should be
accompanied by a flux flow across SIs or sFS parts of the structure.

It is necessary to mention that the results of our investigation may be also
important from the application point of view. Hybrid structures combining
ferromagnetic and superconducting layers became subjects of intensive study
in recent years \cite{RevG, Blamire, Eschrig, LinderRev}. Superconducting
correlations induced into a ferromagnetic region by proximity effect can be
controlled by effective exchange field, leading to a number of practically
important phenomena, such as spin-valve effects \cite{Tagirov, Buzdin2,
Qader, Fominov1, Fominov2, Buzdin3, RevV, Houzet, Linder1, Baek, Soloviev}, which look
rather promising for superconducting electronics \cite{Lu, Tolpygo}. In addition, there is a class of memory devices which operates without performing the magnetization reversal of the ferromagnetic layer \cite{Gold3, Krasnov, Bakurskiy2016, Averin}. The
SIsFS junctions are also considered as possible candidates for memory
elements. \cite{Ruppelt, Larkin, Vernik, Bakurskiy1}. They have a noticeable
advantage compared to standard pseudo spin-valve devices \cite{Baek}. Their $%
I_{C}R_{N}$ product is of the same order as that of the Josephson elements
used in RSFQ logic circuits. In addition, SIsFS junctions can be used as
superconducting transistors \cite{Nevirkovets1, Nevirkovets2}, where the
magnitude and the phase of the order parameter in the middle s-layer are
controlled by spin injection from the F film. The effective magnetic layer F in
these structures can be realized as a composite structure including several
magnetic layers separated by normal or superconductive spacers \cite%
{BakurskiyJETP, HaltermanSFSFS1, HaltermanSFSFS2, Ouassou}.

It is important that the performed investigations of the current
phase relation in SIsFS junctions provide a solid base for
understanding the modes of operation of these transistors and
memory elements.

\textbf{Acknowledgments.} The authors acknowledge helpful discussion with V.V. Ryazanov and E. Goldobin. The developed numerical algorithms and
corresponding calculations in the frame of microscopic model was supported by
the Project No. 15-12-30030 from Russian Science Foundation. This work was
also supported in part by the Ministry of Education and Science of the
Russian Federation in the framework of Increase Competitiveness Program of
NUST "MISiS" (research project K2-2016-051) and grant number MK-5813.2016.2 and by RFBR grants 17-52-560003Iran-a and 16-29-09515-ofi-m.

\appendix

\section{Classification of the states at negative $B$}

Current phase relation with negative sign of the second harmonic amplitude $B
$ in sFS junction can be realized only in the case of more complicated weak
link region. It requires additional inhomogeneity inside F-layer. For
instance, the existence of normal metal areas or step-like geometry of the layer%
\cite{Buzdin, Pugach1, Gold, Bakurskiy3, Gold2, GoldButterfly}.

Generally,  sign change of $B$ leads to a symmetrical transformation
of CPR
\begin{eqnarray}
I_{S}(\varphi ,A,B) &=&-I_{S}(\varphi ,A,-B),  \label{Neg_eq} \\
E(\varphi ,A,B) &=&-E(\varphi ,A,-B).  \label{Neg_eq1}
\end{eqnarray}%
The sign change  of the energy in Eq. \ref{Is4a} significantly influences
the condition of stability (\ref{Stab}). The every stable solution for
positive $B$ becomes unstable after transformation to negative B and vice
versa. This general property determines significant difference between
distributions of states with $B>0$ and $B<0$ on phase plane in Fig.\ref%
{PhaseDiagram2}.

\begin{figure*}[t]
\begin{minipage}[h]{0.3\linewidth}
\center{a)\includegraphics[width=0.95\linewidth]{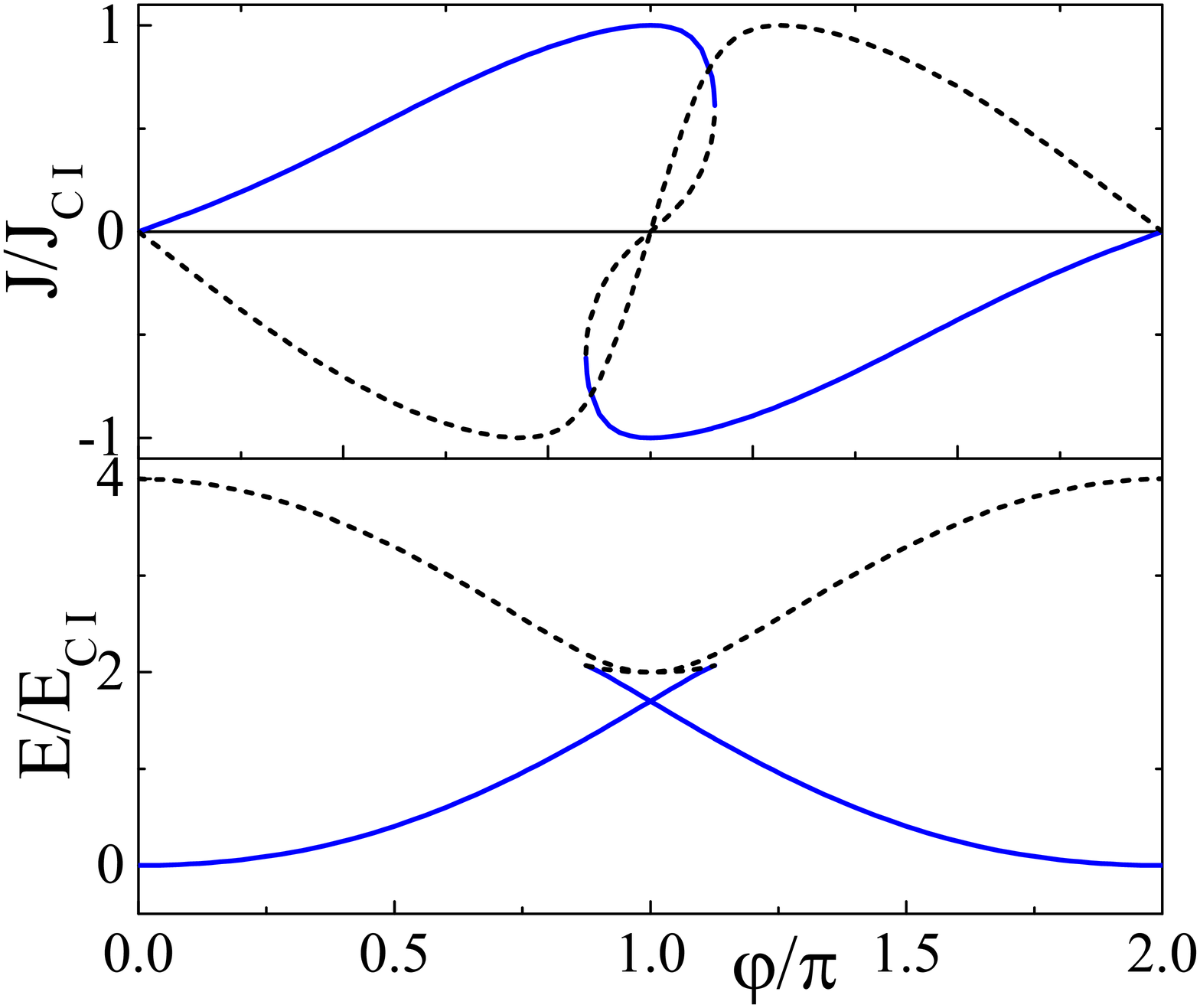}}
\vspace{-5 mm}
\end{minipage}
\hfill
\begin{minipage}[h]{0.3\linewidth}
\center{b)\includegraphics[width=0.95\linewidth]{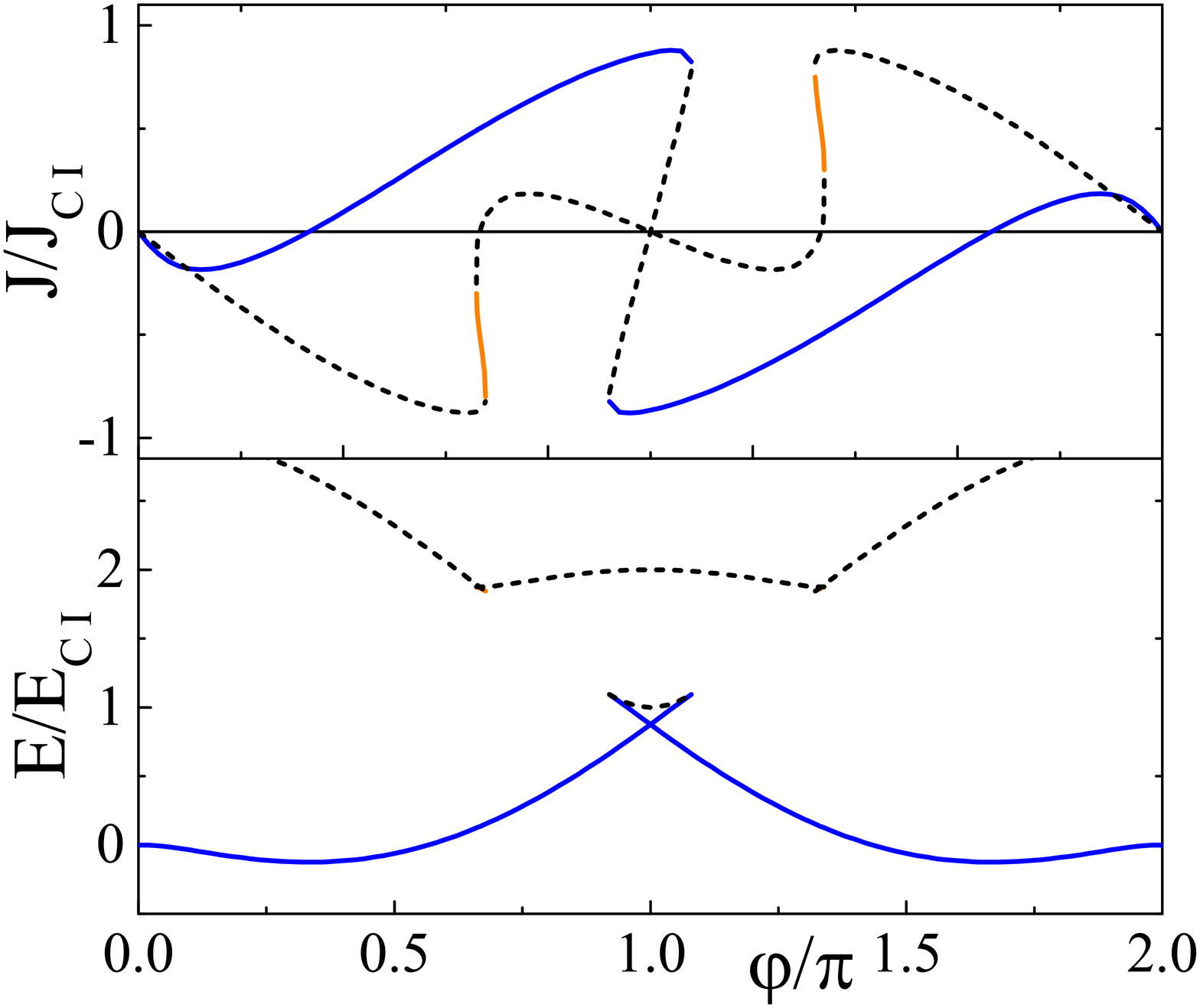}}
\vspace{-5 mm}
\end{minipage}
\hfill
\begin{minipage}[h]{0.3\linewidth}
\center{c)\includegraphics[width=0.95\linewidth]{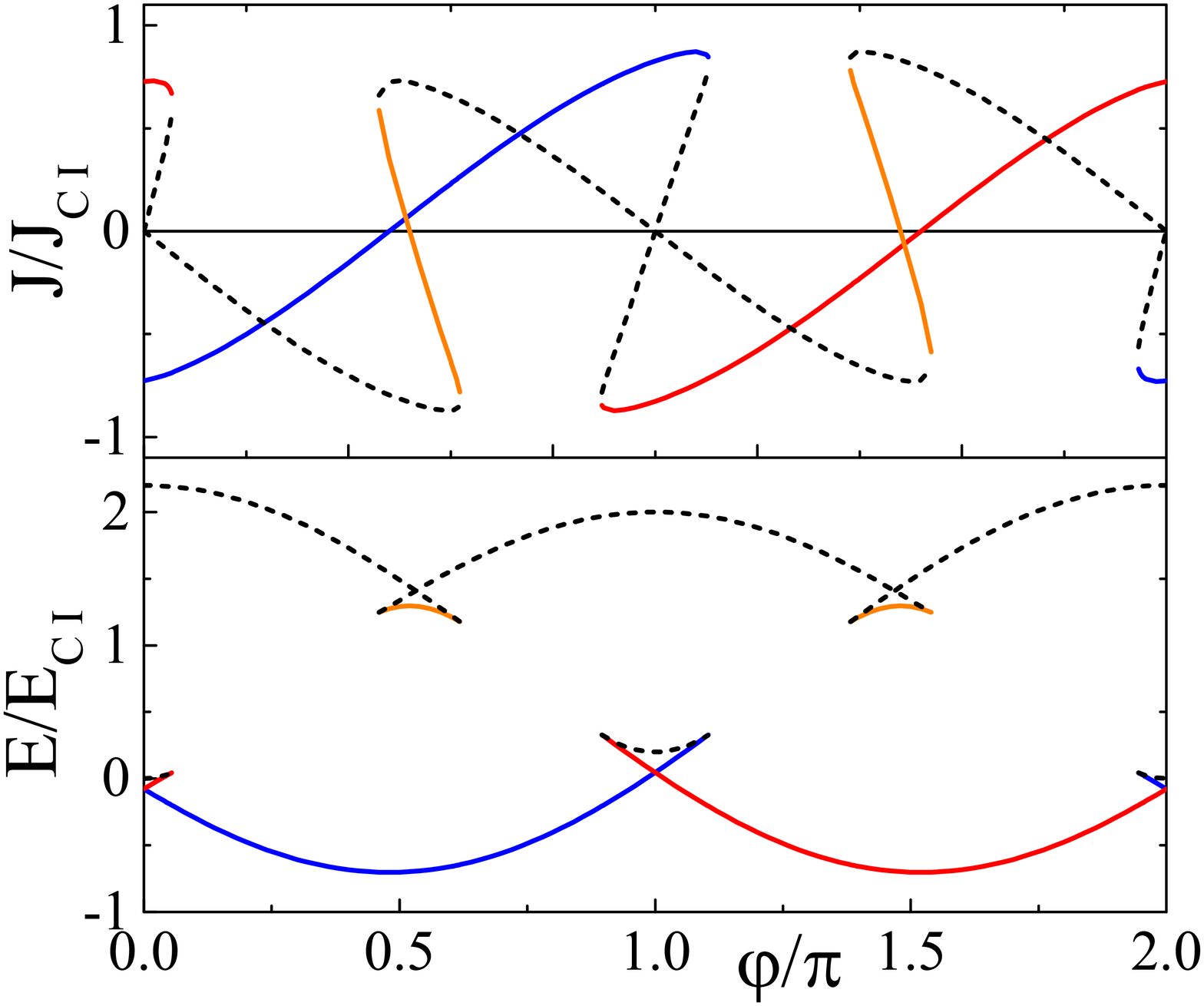}}
\vspace{-2 mm}
\end{minipage}
\vfill
\begin{minipage}[h]{0.3\linewidth}
\center{d)\includegraphics[width=0.95\linewidth]{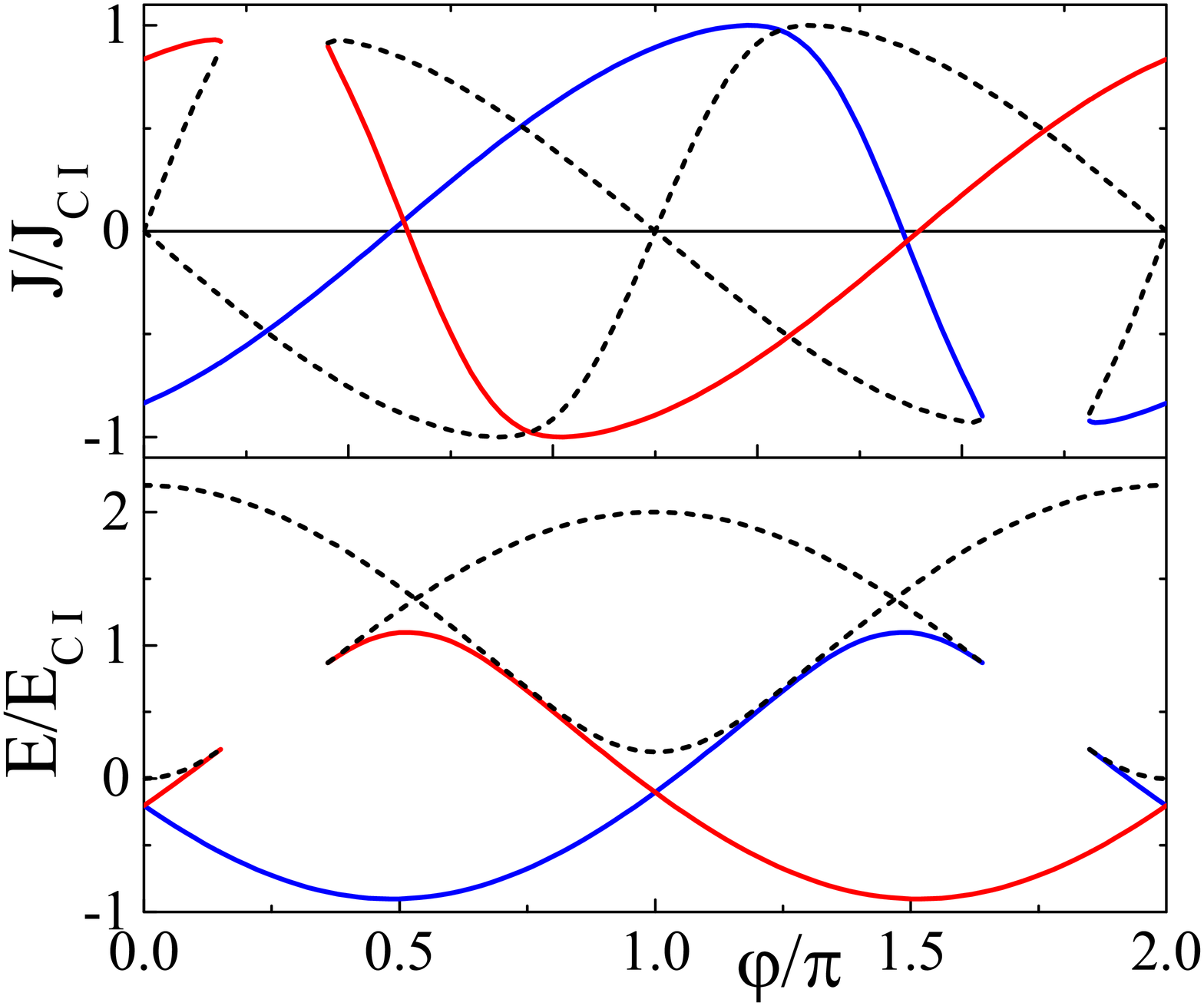}}
\vspace{-2 mm}
\end{minipage}
\hfill
\begin{minipage}[h]{0.3\linewidth}
\center{e)\includegraphics[width=0.95\linewidth]{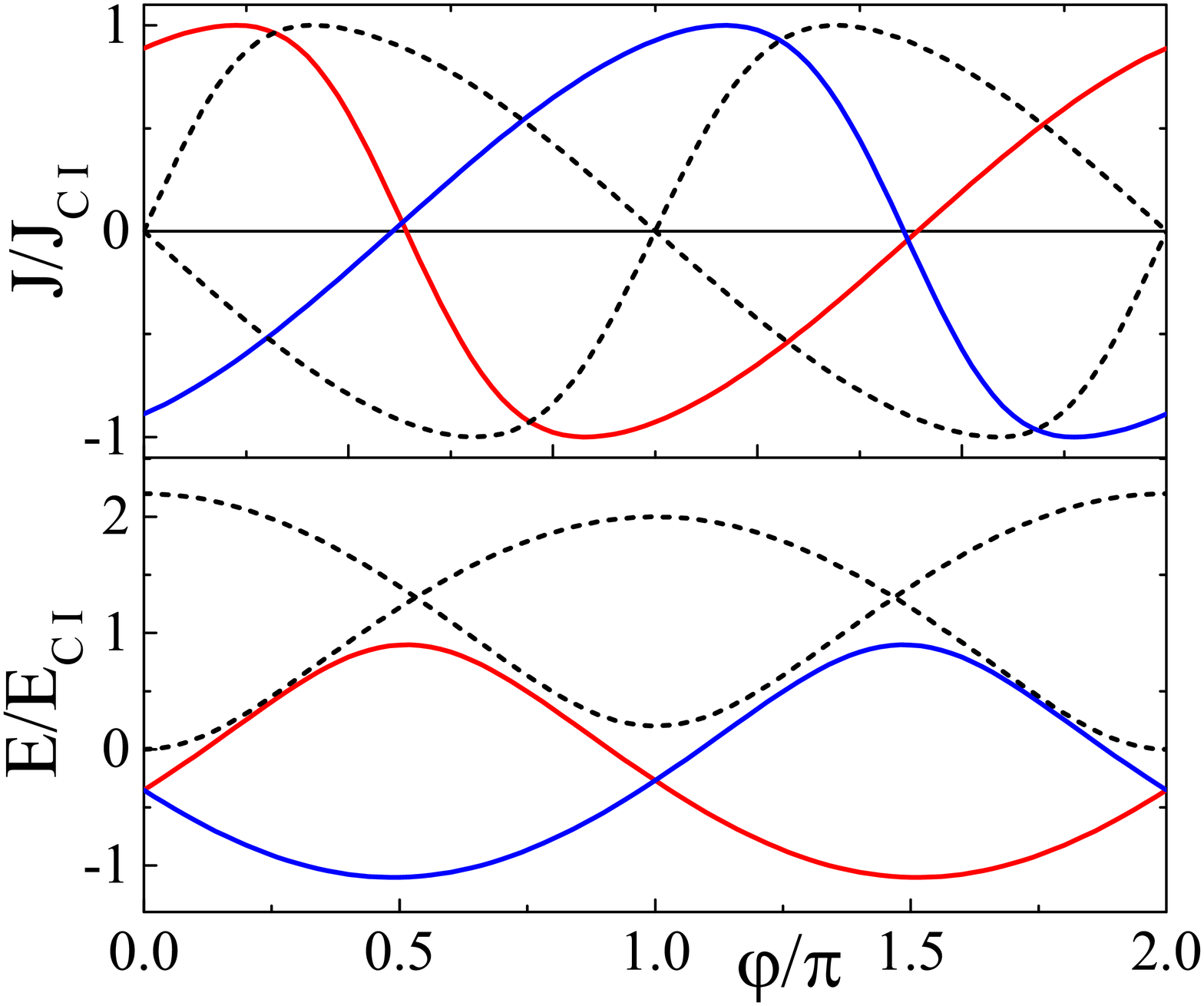}}
\vspace{-2 mm}
\end{minipage}
\caption{(Color Online) The current-phase and the energy-phase relations for
different states of the SIsFS junction with negative amplitudes
of the second harmonic $B<0$: a) 0-ground state with single hysteresis $k=1, m=1$
at $A=1.0$, $B=-0.3$ and b) double-well branch with single hysteresis $k=1,
m=2$ at $A=0.5$, $B=-0.5$; c) hysteretic $\protect\varphi $-state with
double hysteresis $k=2, m=4$ at $A=0.1$, $B=-0.8$; d) $\protect\varphi $
with long hysteresis-state $k=2, m=3$ at $A=0.1$, $B=-1.0$; e) protected $%
\protect\varphi $-state $k=2, m=0$ at $A=0.1$, $B=-1.2$. Solid lines and
dashed lines show stable and unstable states, respectively.}
\label{NegB}
\end{figure*}

Our analysis has shown that, for negative values of amplitude $B$, some new
types of the states may exist (see bottom part of Fig.\ref%
{PhaseDiagram2}). Figure \ref{NegB}a-e demonstrates the main
classes of the current-phase relations existing for $B<0$. Each panel in
Fig. \ref{NegB}a-e gives $I_{S}(\varphi )$ and $E(\varphi )$ dependencies
calculated numerically from equation (\ref{ICeq1}). The dashed black lines
show unstable states. Different colors of the solid lines correspond to
different branches of stable solutions. The filled black circles in the
plane in Fig.\ref{PhaseDiagram2} are the points, at which the CPR presented
in Fig. \ref{NegB}a-e have been calculated in the frame of lumped junctions
model. The number of corresponding pannel in the Fig. \ref{NegB}a-e is
written near the circles.

Typical $I_{S}(\varphi )$ and $E(\varphi )$ curves for the area restricted
by the three lines $B=0.5-0.5|A|$, $B=-0.5+0.5|A|$ and $B=-0.5|A|$ in the $%
A-B$ plane is shown in Fig. \ref{NegB}a. They have been calculated for $%
A=1.0$ and $B=-0.3$. It is seen that $I_{S}(\varphi )$ is a continious function
of $\varphi $ practically for all $\varphi $ except for the area in a vicinity of $%
\varphi =\pi ,$ where the current leap takes place. Figure \ref{NegB}a shows
that there is one ground state in $E(\varphi )$ at $\varphi =0$ and one
hysteresis in $I_{S}(\varphi )$ resulting in $k=1$ and $m=1.$

The considered $A-B$ area provides the first example of the difference in
the SIsFS junctions characteristics between the  cases of positive and
negative  $B$. For $A=1.0$ and  $B=0.3$ there are two hysteresis loops in $%
I_{S}(\varphi )$ relation, while for $A=1.0,$ $B=-0.3$ there is a single
hysteresis loop in CPR. The second hysteresis in $I_{S}(\varphi )$ forms
afterwards, during the further $|B|$ increase.

However, the first effect which appears with $|B|$ \ increase is
transformation of SIsFS structure into a $\varphi -$junction having two
ground states in $E(\varphi )$ dependence, as it is shown in Figs. \ref{NegB}%
b-e.



It is seen that the energy $E(\varphi )$ has two minima at some arbitrary
phases $\varphi =\varphi _{g},$ $\varphi =2\pi -\varphi _{g},$ so that $%
E(\varphi _{g})=E(2\pi -\varphi _{g}).$ This phase $\varphi _{g}$ does not
coincide with both $\varphi =0$ and $\varphi =\pi $ and rapidly saturates at
$\varphi _{g}=\pi /2$ with increasing $|B|$.

The initial stage of $\varphi $-state formation is shown in Fig. \ref{NegB}%
b. It is seen that $I_{S}(\varphi )$ also has a single branch with single
hysteresis $(k=1)$ , but there are two ground states in $E(\varphi )$ curve,
summation over which gives index $m=2$. Fig. \ref{NegB}b demonstrates that the
dependencies are typical for SIsFS contacts if the sFS junction parameters
located inside the area restricted by 
the three lines in $A-B$ plane. They are $B=-0.5-0.5|A|$, $B=-0.5+0.5|A|$
and $B=-0.5|A|$.

After crossing the line $B=-0.5-0.5|A|$ with further $|B|$ increase, the
second hysteresis is nucleating in the current-phase relation in a vicinity
of $\varphi =0.$ Typical $I_{S}(\varphi )$ and $E(\varphi )$ curves for this
range of parameters is demonstrated in Fig. \ref{NegB}c. The calculations
have been done for $A=0.1$ and $B=-0.8$. There is a direct correspondence
between the stable part of the $I_{S}(\varphi )$ curve and the corresponding
ground state in $E(\varphi ).$ This CPR  is characterized by $k=2$ and $%
m=4. $ It is similar to that shown above in Fig. \ref{State1}.

With further $|B|$ growth, the $\varphi$ range of the stable ground solutions increases (blue and red lines on Fig. \ref%
{NegB}c). These branches tend to merge with high energy curves (orange lines
on Fig. \ref{NegB}c). However, this merging doesn't occur simultaneously for
left and right ends of the stable $E(\varphi )$ dependencies. This leads to
formation of the narrow range of parameters, where high energy branches are
connected to the ground branches only from the one end of the stable curve, as it is
shown on the Fig. \ref{NegB}d. The calculations have been done for $A=0.1$
and $B=-1.0$. The corresponding current-phase relation has a long hysteresis,
which provides different indices $m_{i}$ for different ground
states. While we count the number of current jumps during continuous increase
of phase $\varphi $ for the left ground state in Fig. \ref{NegB}d going along the
blue line, the result is the same with later case $m_{1}$=2. However, the index
for the ground state at $\varphi =\pi /2$ on the red line is equal to unity $m_{2}=1$%
. The system jumps on blue line near the phase $\varphi \approx 0.2+2\pi $
and stays on it until $\varphi =\pi /2+2\pi $. Thus the total hysteresis
index $m=3$ is odd for this state, while $k=2$ still the same. Additional
consequence of this property is a dependence of the existing state upon
direction of variation of the phase $\varphi $. If we increase $\varphi $,
the system principally stays on blue line. However, the state on a red line
becomes more probable during the decrease in the phase $\varphi $.

Finally, for $B<-1-2/3|A|$ the system goes in the $\varphi $-state with
protected $I_{S}(\varphi )$ branches, which are characterized
by $k=2$ and $m=0$ (see Fig. \ref{NegB}e). There are two independent $2\pi $
periodical $I_{S}(\varphi )$ curves corresponding to the ground states in the
vicinity $\pi /2$ and $-\pi /2,$ respectively. The calculations have been
done for $A=0.1$ and $B=-1.2$.

\end{document}